\documentclass{vldb}
\usepackage{color}
\usepackage{times}
\usepackage{tikz}
\usepackage{graphicx}
\usepackage{array}
\usepackage{pgfplots}
\usepackage{comment}
\usepackage{breqn}
\usepackage{algorithm}
\usepackage{algpseudocode}
\usepackage{enumitem}
\usepackage{balance}  

\newcommand{\commentout}[1]{%
}
\newcommand{\forlongversion}[1]{#1                                           
}

\newcommand{\coloredcomment}[1]{%
}
\newcommand{\paragrph}[1]{\vspace{5pt}\noindent\textbf{#1}}
\newcommand{\secref}[1]{Section~\ref{#1}}

\newcommand{\figref}[1]{Figure~\ref{#1}}
\newcommand{\tabref}[1]{Table~\ref{#1}}
\newcommand{\eqnref}[1]{Equation~\ref{#1}}
\newcommand{\chapref}[1]{Chapter \ref{#1}}
\newcommand{\G}{G}
\newcommand{\M}{M}
\newcommand{\U}{U'}
\newcommand{\W}{M}
\newcommand{\Q}{Q}
\newcommand{\T}{U}
\newcommand{\R}{G_U}
\newcommand{\bT}{\boldmath{T}}

\newcommand{\refs}{refs}
\newcommand{\I}{I}
\newdef{mydef}{Definition}
\newcommand{\topic}[1]{\vspace{3pt} \noindent \underline{\bf #1}}
\usetikzlibrary{arrows}
\usetikzlibrary{shapes}
\newenvironment{myenumerate}{\begin{enumerate}[labelwidth=\widthof{\ref{last-item}},itemindent=0em,leftmargin=1em,itemsep=0in]}{\end{enumerate}}
\newenvironment{myitemize}{\begin{itemize}[labelwidth=\widthof{\ref{last-item}},itemindent=0em,leftmargin=1em,itemsep=0in]}{\end{itemize}}

\begin{document}


\title{Subgraph Pattern Matching over Uncertain Graphs with Identity Linkage Uncertainty}



%
%
%
%

\numberofauthors{4} 

%
%
\author{
Walaa Eldin Moustafa,~~Angelika Kimmig,~~Amol Deshpande,~~Lise Getoor\\[5pt]
       \affaddr{University of Maryland, College Park, USA} \\[4pt]
       \affaddr{\{walaa, angelika, amol, getoor\}@cs.umd.edu }
}


\maketitle

\begin{abstract}
There is a growing need for methods which can capture uncertainties
and answer queries over graph-structured data.
Two common types of uncertainty are uncertainty
over the attribute values of nodes and uncertainty over the existence
of edges. In this paper, we combine those with 
\emph{identity uncertainty}. Identity uncertainty represents uncertainty over the
mapping from objects mentioned in the data, or \emph{references}, to
the underlying 
real-world \emph{entities}. 
We propose the notion of a \emph{probabilistic entity graph} (PEG),
a probabilistic graph model that defines a distribution
over possible graphs at the entity level. The model takes into account node attribute uncertainty, edge existence
uncertainty, and identity uncertainty, and thus enables us to systematically reason about
all three types of uncertainties in a uniform manner. We introduce a general framework for 
constructing a PEG given uncertain data at the reference level and 
develop highly efficient algorithms to 
answer subgraph
pattern matching queries in this setting.
Our algorithms are based on two novel ideas: \emph{context-aware path indexing} and \emph{reduction by join-candidates}, 
which drastically reduce the query search space. 
A comprehensive experimental evaluation shows that our approach
outperforms baseline implementations by orders of magnitude.

\end{abstract}

\section{Introduction}

The ability to reason about relationships between objects or entities in the presence of uncertainty is crucial  in a variety of application domains, 
such as 
online social networks, the Web, communication networks, bioinformatics and financial data management. 
As data in these domains is naturally modeled as {\em graphs}, a range of scalable algorithms and indexing techniques
for analyzing and querying graphs has been developed recently, 
e.g., \cite{yan:sigmod04,cheng:sigmod07,he:sigmod08,zou:vldb09,fan:vldb10}. With the exception of some recent work, e.g., \cite{chen:tkde10,jin:kdd11,lian:sigmod11,papapetrou:edbt11,yuan:vldb12}, most prior work 
ignores the uncertainties that are often  inherent in the data. 

\commentout{
The reasons for uncertainty are manyfold. Uncertainty can be used to summarize common data quality issues such as missing or inaccurate information. The uncertainties themselves may come from an information extraction system which returns a confidence associated with an extracted fact \cite{suchanek:www07,carlson:aaai10}. The probabilities may be the output of some machine learning system, or a domain expert may be able to specify them. Another type of uncertainty arises when integrating information from multiple sources. The same object may appear under different names or forms in each source, leading to duplicate copies of the object. Although there is much work on entity resolution, e.g., \cite{benjelloun:vldb08,arasu:icde09,getoor:pvldb12}, often times the available  information is not sufficient to resolve the ambiguities with certainty.
}
Such networks are often created by applying well-established statistical and probabilistic methods for tasks such as  information extraction, information
integration, predictive analysis, node classification, link prediction, entity
resolution, and community detection. Applying these methods to observed data naturally produces
probabilistic graphs with different types of uncertainties, and many
of these approaches quantify this uncertainty. For instance,
information extraction systems often return a  confidence associated
with an extracted fact \cite{suchanek:www07,carlson:aaai10}. Another
type of uncertainty arises when integrating information from multiple
sources. The same object may appear under different names or forms in
each source, leading to duplicate copies of the object. Although there
is much work on entity resolution, e.g.,
\cite{benjelloun:vldb08,arasu:icde09,getoor:pvldb12}, the
available  information is often not sufficient to resolve the ambiguities
with certainty. 


In this work, we abstract from the concrete source of uncertainty and
propose a general probabilistic graph model  that combines three common  types of uncertainty. Specifically, we consider: 1) uncertainty about the attribute values of nodes (i.e., \emph{attribute value uncertainty}), 2) uncertainty about whether particular edges exist (i.e., \emph{edge existence 
uncertainty}), and 3)  \emph{identity uncertainty}, that is, uncertainty about whether each real world entity is represented by one or multiple objects or identifiers  in the data.

In addition, we develop techniques for efficiently answering subgraph
pattern queries over such uncertain graphs. 
We show that our model defines a probability distribution over possible graphs describing entities, their labels and relations. 
We then introduce techniques to find all matches of a subgraph pattern that have a probability above a given threshold. 
Answering subgraph pattern matching queries is NP-hard on non-probabilistic graphs. It becomes even harder when adding uncertainty,
especially identity uncertainty, making the problem \#P-complete. 
Nonetheless, we propose and systematically explore a range of novel techniques to prune the search space and effectively perform subgraph pattern matching over large-scale uncertain graphs.

To summarize, we make the following contributions:
\begin{myitemize}
\item We introduce \emph{probabilistic entity graphs}, a general uncertain graph model that captures attribute, edge and identity uncertainties.
\item We define the semantics of probabilistic entity graphs as a probability distribution over possible entity graphs.
\item We develop scalable algorithms to answer subgraph pattern matching queries over such uncertain graph data, based on \emph{query path decomposition}.
\item We present a novel graph indexing method,
  \emph{context-aware path indexing}, to capture
  information about the graph paths, their surrounding structures, and their probabilities, enabling efficient retrieval of candidate matches.
\item We propose \emph{reduction by join-candidates}, an algorithm that efficiently prunes candidate answers by progressively propagating structural and probabilistic information between the candidates.
\item We demonstrate that our approaches can evaluate complex queries over graphs with millions of nodes and edges in seconds, outperforming a baseline implementation by orders of magnitude.
\end{myitemize}

\section{Motivating Example}

\begin{figure}
\begin{center}
\begin{tabular}{c c c c}
\includegraphics[scale= 0.15]{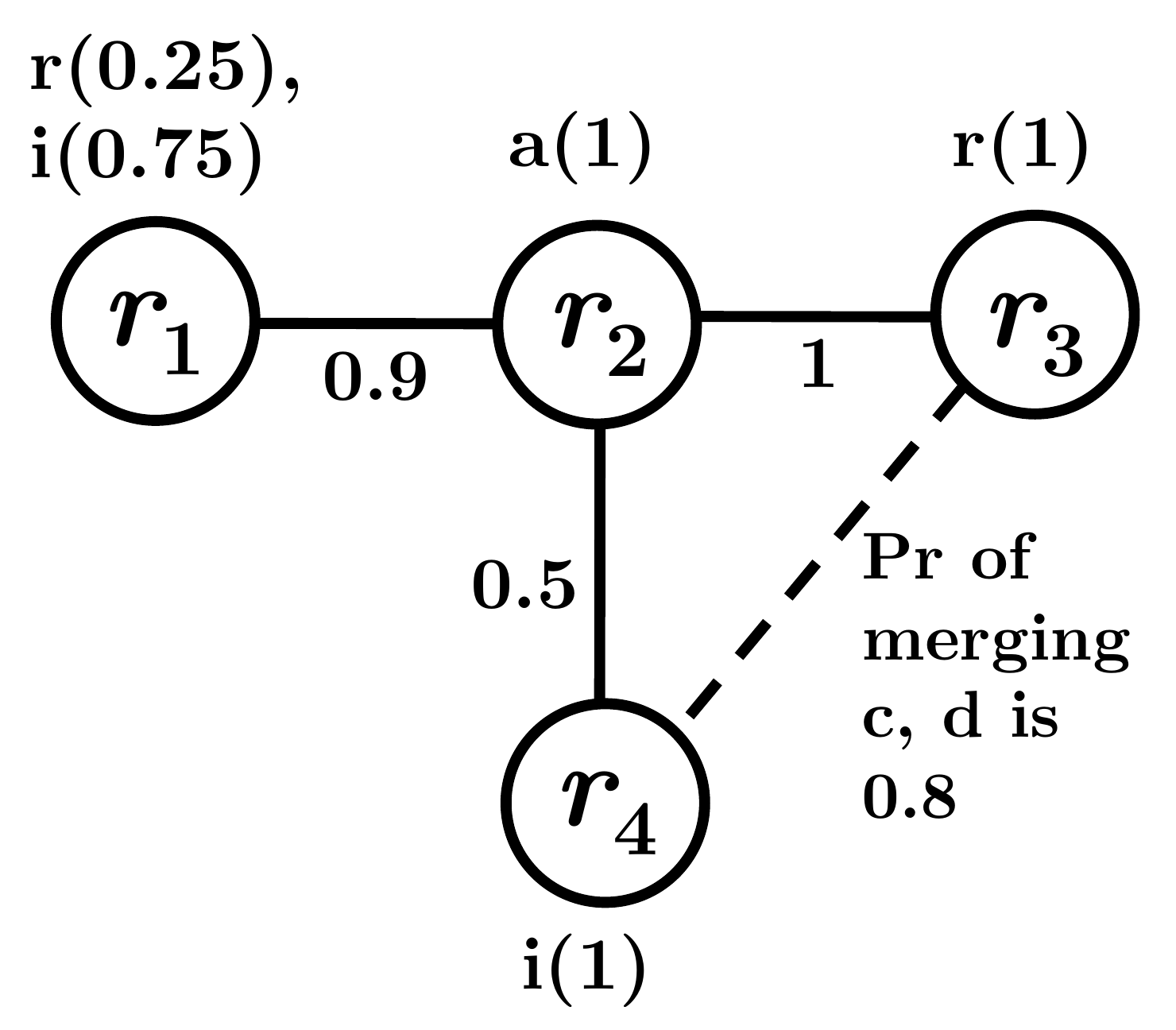}
&
\includegraphics[scale= 0.15]{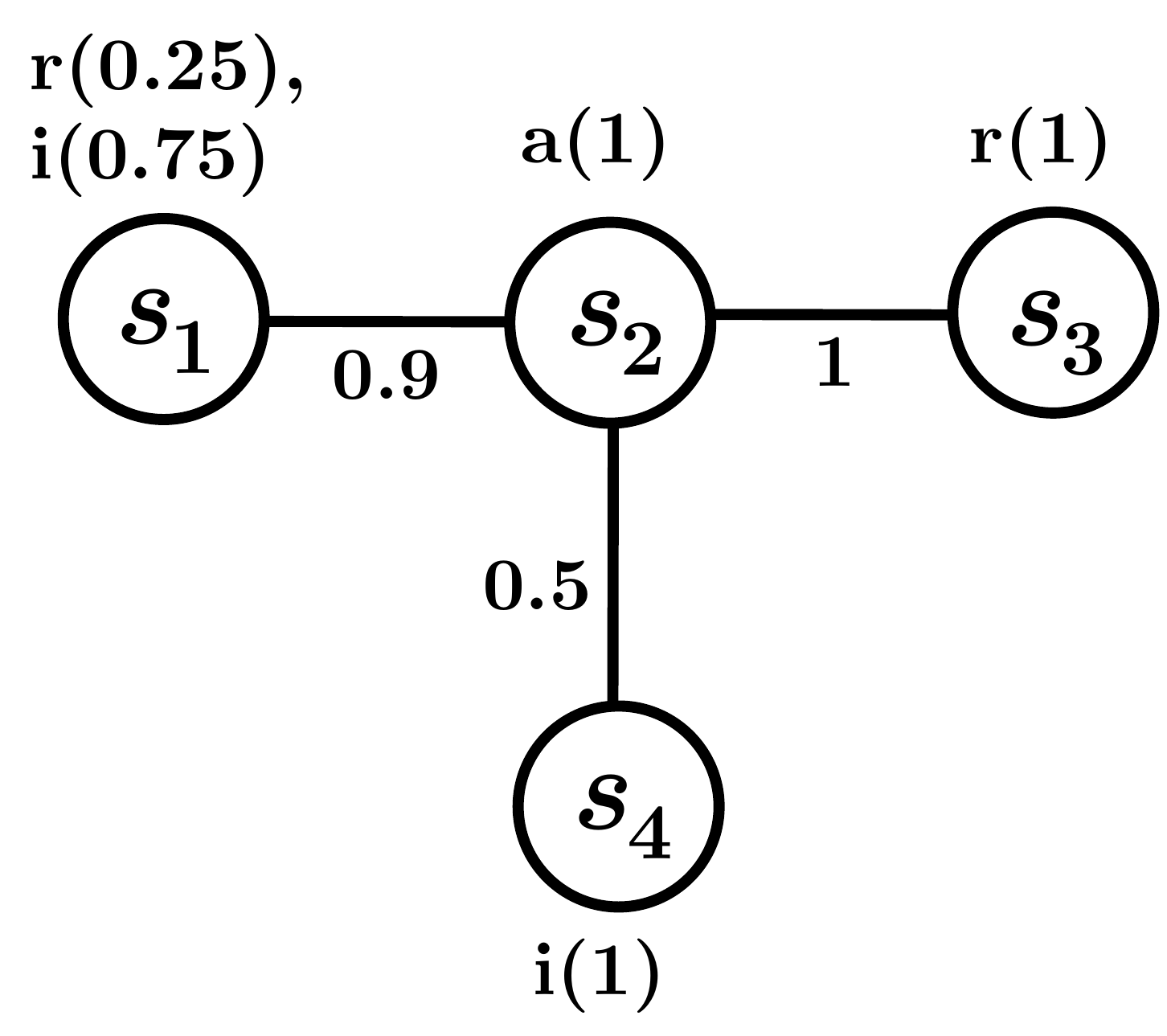}
&
\includegraphics[scale= 0.15]{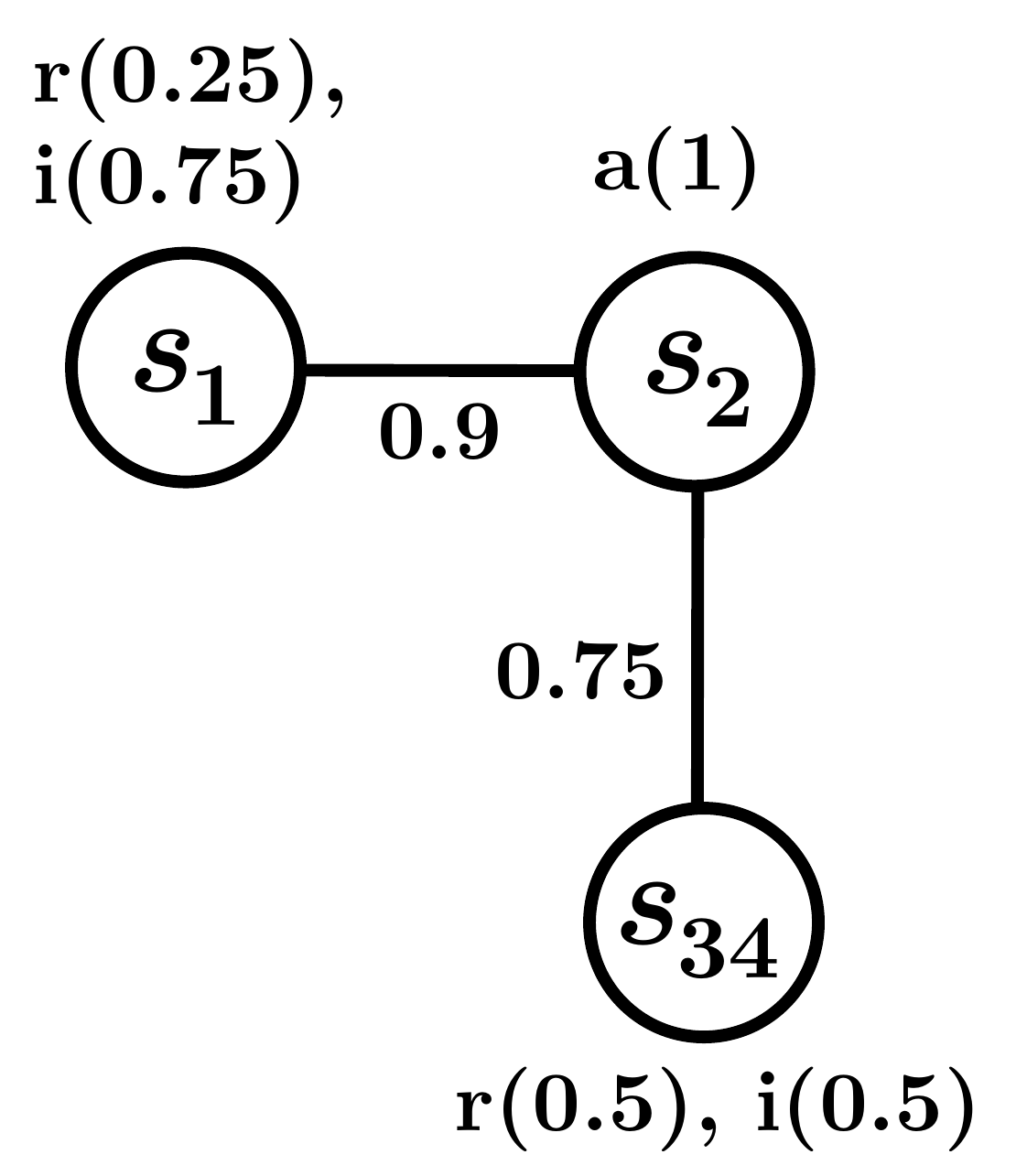}
&
\parbox{0.2in}{\vspace{-0.72in}\includegraphics[scale=0.15]{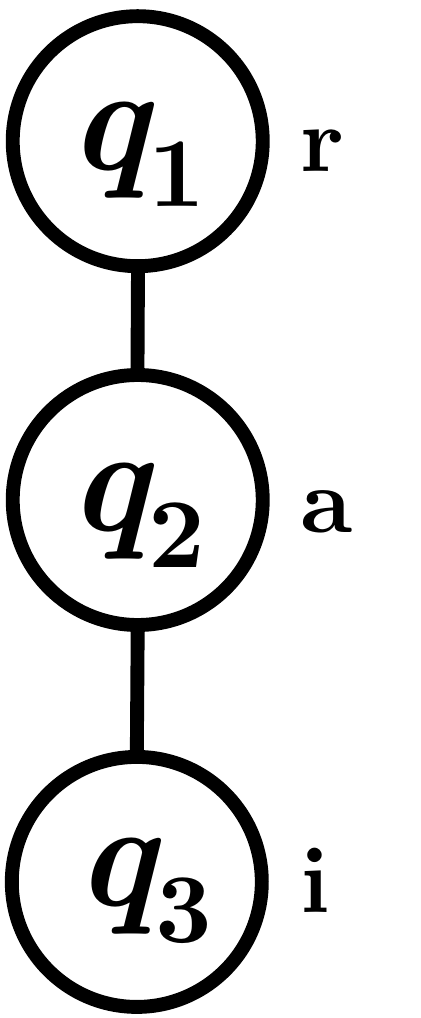}}
\\

(a) & (b) & (c) & (d)\\
\end{tabular}
\end{center}
\vspace{-15pt}
\caption{(a) Reference-level network, (b), (c) the two possible entity graphs, (d) a query graph}
\vspace{-10pt}
\label{fig:general-example}
\end{figure}

Consider a system to help organizations find experts in different domains. The system integrates information about experts and their affiliations from multiple sources. Assume three sources: an online professional network (e.g., LinkedIn), an online social network (e.g., Facebook), and personal webpages or blogs. The system makes used of the experts' names, their affiliations (specifically, \texttt{Academia} (a), \texttt{Research Lab} (r), or \texttt{Industry} (i)), and relationships between experts. 
\figref{fig:general-example} illustrates a small example, where we omit names for clarity. We use the term \emph{reference} to denote the \emph{observed objects}, which in this example are strings encoding names, 
while we use the term \emph{entity} to refer to \emph{real-world objects}, that is, the experts in our case. A real-world object may thus correspond to a collection of references, as names may be abbreviated, misspelled, etc. In \figref{fig:general-example}(a), nodes represent references, letters inside nodes represent reference IDs, and letters outside nodes represent \emph{labels}, that is, affiliations, along with their probabilities in parentheses.  Consider node $r_1$, extracted from a personal webpage. Suppose that a text analysis method suggests that the name is ``Gerald Maya'' and the affiliation is \texttt{industry} with probability $0.75$ and a \texttt{Research Lab} with probability $0.25$. Nodes $r_2$ and $r_3$ are extracted from an online professional network, with name ``Becky Castor'' and an \texttt{Academia} affiliation, and the name ``Christopher Tucker'' and a \texttt{Research Lab} affiliation, respectively. Finally, node $r_4$ is extracted from an online social network, with the name ``Chris Tucker'' and an \texttt{Industry} affiliation. Furthermore, relationships between the individuals are extracted (represented as edges in the figure) and are associated with probabilities that reflect the likelihood of the relationship's existence.  These probabilities can be calculated based on whatever information or signals is available from these online resources, such as the number of common connections or shared attributes between them. Since ``Christopher Tucker'' and ``Chris Tucker'' seem to be the same person based on name similarity, we put them together in the same \emph{reference set} to indicate that these two references may refer to the same entity (depicted as a dashed line in the figure).  To quantify identity uncertainty, which is the uncertainty of having multiple references referring to the same real-world entity, we assign this reference set a probability of $0.8$, denoting the likelihood that the elements in the set correspond to a single real-world entity. 

Figures \ref{fig:general-example}(b) and (c) illustrate the two possible sets of entities with their labels and relations for the example reference network shown in \figref{fig:general-example}(a), where the letters inside the nodes represent entity IDs. \figref{fig:general-example}(b) depicts the entity
graph in which $r_3$ and $r_4$ remain unmerged, i.e., assumed to be separate real-world entities, ($pr = 0.2$), 
and \figref{fig:general-example}(c) depicts the case
where they are merged, i.e., assumed to be the same real-world entities, ($pr = 0.8$) 
to form a new node $s_{34}$ with its own label and edge probability distributions.
Going from a set of references to an entity requires merging the information associated with the references, that is, their labels and the relationships they participate in. 
In this example, we simply average the probability distributions. 
Since
$r_3$ has label~$r$ and $r_4$ has label~$i$,
we assign a label distribution of $r(0.5), i(0.5)$ to entity $s_{34}$.
Similarly, 
$s_{34}$ has an edge to $s_2$ with $pr = 0.75$ (average of $r_3$'s edge with $pr=1$ and
$r_4$'s edge with $pr=0.5$).

Clearly, we want to specify queries to our information system at the level of entities rather than references.
In this work, we focus on subgraph pattern matching queries, perhaps the most widely used and studied class of queries over graphs. \figref{fig:general-example}(d) depicts a query
which asks for all paths of length 2 over nodes labeled $(r, a, i)$.
In addition to the query
graph, a query specifies a minimum threshold $\alpha$,
which we set to $0.25$ in this example, to indicate that only matches with
probability larger than $\alpha$ should be returned.
In this simple case, we can answer our query by examining all possible matches. 
In the 
entity graph in \figref{fig:general-example}(b), with $r_3$ and $r_4$ unmerged, the nodes $(s_3, s_2, s_4)$ form a path with the required
labels. The probability of that path is computed by
multiplying together the three node label probabilities (1, 1, 1), the
two edges probabilities (1, 0.5), and the probability that the nodes $r_3$
and $r_4$ are {\em not} merged (0.2); the resulting probability of the match is 0.1, which is below
our cutoff of $0.25$.
There are two more potential matches, 
$(s_1,s_2,s_4)$ and $(s_3,s_2,s_1)$, but neither of them satisfies the
minimum threshold constraint. In the second 
entity graph in \figref{fig:general-example}(c), there
are two potential matches for the query: $(s_1,s_2,s_{34})$ and $(s_{34},s_2,s_1)$. The
probability of $(s_1,s_2,s_{34})$ being a match to the query is $0.084$, which does not meet our threshold,
whereas the probability of $(s_{34},s_2,s_1)$ is $0.253$.
Therefore, $(s_{34},s_2,s_1)$ is the only answer to our query. 
Clearly, such an exhaustive approach is infeasible in practice for
larger graphs. In this work, we therefore develop a scalable approach to answer subgraph pattern matching queries in this setting.

\section{Uncertain Graph Modeling}
\label{sec:modeling}
We now discuss our formal model for the types of uncertainties arising in situations as described in the example above, where we are given information about \emph{references}, or mentions of objects, but are interested in queries about \emph{entities}, or the objects themselves. 
We introduce \emph{probabilistic entity graphs}, which define a probability distribution over graphs describing entities, their labels, and links between them. The key challenge here is that references induce 
constraints on which entity nodes can co-occur in the same graph, as each graph structure corresponds to one possible way of assigning references to existing entities. 
To deal with these dependencies, we represent our probability distribution as a  
\emph{probabilistic graphical model} (PGM)
\cite{koller:pgmbook09}. After a quick summary of  the necessary basics, 
we introduce the notion of a \emph{probabilistic graph description} (PGD), and show how the PGD in turn defines a probabilistic entity graph.
We first focus on the basic case, where distributions over labels and links are all independent, and then show how additional dependencies can directly be introduced. \forlongversion{ Notations for this Section are shown in \tabref{tab:notations}.}

\forlongversion{
\begin{table}
\begin{tabular}{|p{0.75in}| p{2.3in}|}
\hline
Notation & Definition\\
\hline
$\Sigma$ & Set of labels\\
$R$ & Set of references\\
$S$ & Set of sets of references\\
$r$ & Reference in $R$\\
$s$ & A set in $S$ representing a potential real-world entity\\
$r.x$ & Random variable representing the reference's label\\
$(r_1,r_2)$ & Edge in $R \times R$\\
$(r_1,r_2).x$ & Random variable representing the edge's existence\\
$s.x, s.n$ & Random variables representing the existence of an entity (used interchangeably in the contexts of PGD and PEG, respectively)\\
$s.l$ & Random variable representing the entity's label\\
$(s_1,s_2).e$ & Random variable representing the existence of edge between entities $s_1$ and $s_2$\\
$S.\boldsymbol{n}$ & Shorthand for $s_1.n=n_1, \ldots , s_{|S|}.n=n_{|S|}$\\
$S^r=\{s_1, \ldots, s_k\}$ & Subset of $S$ that contains all sets that contain $r$, i.e., $\{s\in S|r\in s\}$\\
$v$ & Entity graph node\\
$e$ & Entity graph edge\\
$v.n, v.l$ & Entity graph random variables for node's existence, and node's label, respectively\\
\hline
\end{tabular}
\caption{Notations used in \secref{sec:modeling}}
\label{tab:notations}
\end{table}
}

A PGM $\mathcal{P} = \langle \mathcal{V} , \mathcal{F} \rangle$ defines a joint probability distribution over its random variables $\mathcal{V}$ via its set of factors~$\mathcal{F}$. Each factor $f$ is defined over a subset $\mathcal{V}_f$ of $\mathcal{V}$ and represents a dependency between those random variables. Given a complete joint assignment $\boldsymbol{v} \in Dom(\mathcal{V})$ to the variables in $\mathcal{V}$, the joint distribution is defined by $Pr(\boldsymbol{v}) =\frac{1}{\mathcal{Z}}\prod_{f \in \mathcal{F}}f(\boldsymbol{v}_f)$, where $\boldsymbol{v}_f$ denotes the assignments restricted to the arguments $\mathcal{V}_f$ of $f$ and $\mathcal{Z}= \sum_{\boldsymbol{v}'\in Dom(\mathcal{V})}\prod_{f \in \mathcal{F}}f(\boldsymbol{v}'_f)$ is a normalization constant referred to as the \emph{partition function}. 
The independencies in the distribution defined by a PGM are represented graphically in its \emph{Markov network}, which 
contains one node for each random variable, and an edge between a pair of random variables if and only if the two variables co-occur in some factor. Each connected component in the Markov network corresponds to a part of the model that is \emph{independent} from the rest. We can thus  compute the normalized probability for each connected component separately and multiply them together to obtain the full joint distribution.   

As a first step towards our probabilistic model, we now introduce random variables for labels of references ($r.x$), existence of edges between pairs of references ($e.x$), and existence of an entity corresponding to a set of references ($s.x$). We further specify a probability distribution over each such random variable. 
\begin{mydef}
\textbf{Probabilistic Graph Description:}  
A probabilistic graph description (PGD) is a tuple $D = (R, S, \Sigma, P, m^{\Sigma}, m^{\{T,F\}})$, where
$R$ is a set of references, $S$ is a set of subsets of~$R$ including at least all singleton subsets, 
$\Sigma$ is a set of labels, and:
\begin{myitemize}
\item $P$ is a set of probability distributions containing (1) for each $r\in R$, a probability distribution $p^r(r.\boldsymbol{x})$ over a random variable $r.x$ with values from $\Sigma$, (2)
for each $(r_1,r_2)\in R\times R$, a probability distribution $p^{(r_1,r_2)}((r_1,r_2).\boldsymbol{x})$ over a random variable $(r_1,r_2).x$ with values from $\{T,F\}$, and (3) for  each $s\in S$, a probability distribution $p^s(s.\boldsymbol{x})$ over a random variable $s.x$ with values from $\{T,F\}$.
\item  The merge functions 
$m^{\Sigma}$ and $m^{\{T,F\}}$ transform a  set of probability distributions over random variables with values in $\Sigma$ and $\{T,F\}$, respectively,  into a single such distribution.
\end{myitemize}
\end{mydef}

For example, in \figref{fig:general-example}(a), $R=\{r_1,\ldots,r_4\}$, $S=\{\{r_1\},\{r_2\},\\ \{r_3\},\{r_4\}, \{r_3,r_4\}\}$, $\Sigma=\{a,r,i\}$, $P$ includes the given probability distributions, and
finally, both $m^{\Sigma}$ and $m^{\{T,F\}}$ simply {\em average} the input probability distributions (this is also the merge function we use in our experimental evaluation).

A PGD thus specifies the set of observed references $R$ together with their possible labels as well as probabilities for the existence of edges between two references. Each set in $S$ corresponds to a potential entity and contains all references to that entity. 
The PGD specifies independent probability distributions for the existence of such entities. The merge functions are used to compute new probability distributions after merging two or more references into a single entity. Different merge functions are appropriate in different settings. 
Aside from {\em average} described above,
another example of a merge function for $m^{T,F}$ is {\em disjunct}, where the output probability distribution is the disjunction of the input distributions. 

In the next step of our model construction, the probabilistic entity graph combines these independent probability distributions into a graphical model that encodes the dependencies between entities induced by shared references and combines the distributions over labels and edges using the merge functions provided by the PGD. 


\begin{mydef}
\textbf{Probabilistic Entity Graph:}  
For a given PGD~$D$, the probabilistic entity graph (PEG) $U$ is a graphical model with set of random variables $\mathcal{V} = \{s.n|s\in S\}\cup\{s.l|s\in S\}\cup\{e.e|e\in S\times S\}$ and set of factors $\mathcal{F}$ defined as follows. For each $r\in R$ with $S^r = \{s_1, \ldots, s_k\}=\{s\in S|r\in s\}$, 
$\mathcal{F}$ contains a node existence factor  
\begin{dmath}
\nonumber
{f^{N}(s_1.n=v_1,\ldots, s_k.n=v_k)}
=
\begin{cases}
p^s(s_i.x=T) & \text{if }v_i=T \text{ and }v_j=F\text{ for all }j\neq i\\
0 & \text{otherwise.}
\end{cases}
\end{dmath}
For each $s\in S$, $\mathcal{F}$ contains a node label factor \coloredcomment{\color{red} in next two equations, parens are confusing; the idea is that we first apply the merge function to the set, which results in a function, and that function's argument is the rv; any idea how to simplify notation?}
\begin{equation}\label{eq:mergelabels}
Pr(s.\boldsymbol{l}) = \left[m^{\Sigma}(\{p^r| r\in s\})\right](s.\boldsymbol{l})
\end{equation}
For each $(s_1,s_2)\in S\times S$, $\mathcal{F}$ contains an edge existence factor
\begin{equation}\label{eq:mergeedges}
Pr((s_1,s_2).\boldsymbol{e}) = \left[m^{\{T,F\}}(\{p^{(r_1,r_2)}|r_i\in s_i\})\right]((s_1,s_2).\boldsymbol{e})
\end{equation}
\end{mydef}

Identity uncertainty is modeled by the node existence factors $(f^{N}(s_1.n=v_1,\ldots, s_k.n=v_k))$, which ensure that all assignments where two entity nodes share a reference have zero probability. The node label factors ($Pr(s.\boldsymbol{l})$) are probability distributions obtained by aggregating the label probability distributions of all references in the underlying set $s$ via the node label merge function. In the same way, the edge existence factors ($Pr((s_1,s_2).\boldsymbol{e})$) are probability distributions obtained by aggregating the edge existence probability distributions of all pairs of references from the underlying sets via the edge existence merge function. 

\vspace{8pt}
\noindent
\textbf{Exploiting Independence:} Writing out the probability distribution defined by the PEG, we have
\begin{dmath}
\mbox{~}\hspace{-8pt}Pr(S. \boldsymbol{n},S. \boldsymbol{l},(S\times S).\boldsymbol{e})
= \frac{1}{\mathcal{Z}} \cdot \prod_{r\in R} f^{N}(S^r.\boldsymbol{n}) \cdot \prod_{s\in S} Pr(s.\boldsymbol{l}) \\ \cdot \prod_{(s_1,s_2)\in S\times S} Pr((s_1,s_2).\boldsymbol{e})
\label{eq:pgm}
\end{dmath}
We use shorthand notation for assignments to sets of random variables, e.g., $S.\boldsymbol{n}$ for $s_1.n=n_1, \ldots , s_{|S|}.n=n_{|S|}$.  
The partition function $\mathcal{Z}$ is the sum of the factor product over all variable assignments. As all node label and edge existence factors are probability distributions  independent of all other factors, Eq.~\ref{eq:pgm} is equivalent to 
\begin{dmath}
Pr(S.\boldsymbol{n},S.\boldsymbol{l},(S\times S).\boldsymbol{e})
= Pr(S.\boldsymbol{n}) \cdot \prod_{s\in S} Pr(s.\boldsymbol{l}) \cdot \prod_{(s_1,s_2)\in S\times S} Pr((s_1,s_2).\boldsymbol{e})
\label{eq:pgmpr}
\end{dmath} 
where $Pr(S.\boldsymbol{n})$ is the normalized product of all node existence factors, that is, the partition function $\mathcal{Z}_n$ is with respect to those factors only:
\begin{equation}
Pr(S.\boldsymbol{n}) = \frac{1}{\mathcal{Z}_n}\prod_{r\in R}  f^{N}(S^r.\boldsymbol{n})
\end{equation}
It is often possible to further decompose this function, taking into account the independencies encoded in the Markov network. Let $\mathcal{C}(S.n)$ be the partitioning of the set of random variables $S.n$ induced by  the connected components of the Markov network, that is, each element of $\mathcal{C}(S.n)$ contains all random variables participating in one such component. We can then rewrite the above equation as
\begin{align}
Pr(S.\boldsymbol{n}) & = \prod_{S_i.n\in\mathcal{C}(S.n)} \frac{1}{\mathcal{Z}_{n_i}}\prod_{r\in R \wedge S^r\subseteq S_i}  f^{N}(S^r.\boldsymbol{n})\nonumber\\
&=  \prod_{S_i.n\in\mathcal{C}(S.n)}Pr(S_i.\boldsymbol{n})
\label{eq:splitnodeprob}
\end{align}
where the partition function $\mathcal{Z}_{n_i}$ normalizes over all assignments for random variables in $S_i.n$.

\vspace{4pt}
\noindent\textbf{Distribution over Graphs.} Clearly, not all assignments to random variables in the model above directly correspond to legal graphs. 
We now show how to obtain the final distribution over labeled graphs.
The set of possible world graphs $PW(U)$ of a PEG $U$ consists of those 
 graphs $W=(V,E,l(.))$  where $V$ is a set of entity nodes corresponding to reference sets from~$S$ (merged into a single entity), $E\subseteq V\times V$ is a set of edges between them, and the label function $l: V\rightarrow \Sigma$ labels these nodes with elements of $\Sigma$. Slightly abusing notation, we often identify a graph node $v\in V$ with the corresponding set of references $s\in S$, and use both notations interchangeably. This allows us to treat $V$ as a subset of $S$ and thus simplify notation. 
Each possible world graph~$W$ induces a partial value assignment $(S.\boldsymbol{n}^W,V.\boldsymbol{l}^W,(V\times V).\boldsymbol{e}^W)$ to the random variables in the graphical model as follows. For each $s\in V$, we have $s.n^W=T$, and for each $s\in S\setminus V$,  we have  $s.n^W=F$, that is, values of node existence variables mirror the (non-)existence of nodes in $W$. For each $s\in V$, we have $s.l^W=l(s)$, that is, for all existing nodes, values of node label random variables mirror the labels in $W$, and all other node label random variables remain unassigned. For all $(s_1,s_2)\in E$, we have $(s_1,s_2).e^W=T$, and for all $(s_1,s_2)\in (V\times V) \setminus E$, we have $(s_1,s_2).e^W=F$, that is, for all pairs of existing nodes, edge existence variables mirror the (non-)existence of edges in the graph, and all other edge existence random variables remain unassigned. The probability of~$W$ is now obtained based on \eqnref{eq:pgmpr} by marginalizing over all unassigned variables. As those all appear in independent factors only, we get
\begin{dmath}
Pr((V,E,l(.))) = Pr(S.\boldsymbol{n}^W) \cdot \prod_{v\in V} Pr({v.l=l(v)}) \cdot \prod_{(s_1,s_2)\in E} Pr({(s_1,s_2).e=T}) \cdot \prod_{(s_1,s_2)\in (V\times V) \setminus E} Pr({(s_1,s_2).e=F})\label{eq:possworldprob}
\end{dmath}
As every full assignment to the variables in the graphical model contributes to exactly one graph's probability, this defines a probability distribution over possible world graphs. 

\vspace{5pt}
\noindent\textbf{Introducing Correlations.} So far, 
while we allow arbitrary dependencies among
node existence probabilities, for ease of exposition, we have assumed that
 node label and edge existence probabilities are independent. The probabilistic model
can easily accomodate dependencies between attributes and edges, as long as the
dependencies remain \emph{acyclic} \cite{getoor:jmlr02}.

As a simple yet useful example of the types of correlations that can
be accommodated, in this work, we model correlations between edge
existence and attribute values given by labels. More specifically, we consider 
the case where the existence of each edge depends on the labels of the
two nodes it connects. 
This can be achieved by replacing the edge existence probabilities
$p^{(r_1,r_2)}((r_1,r_2). \boldsymbol{x})$ in the PGD by conditional
probabilities
$p^{(r_1,r_2)}((r_1,r_2). \boldsymbol{x_0}|r_1. \boldsymbol{x_1},r_2. \boldsymbol{x_2})$,
that is, the probability of an edge existing between two references
depends on the labels of these references. This means that the edge
existence factors in the PEG now include three random variables, with
values
$Pr((s_1,s_2).\boldsymbol{e}|s_1.\boldsymbol{l_1},s_2.\boldsymbol{l_2})$
obtained by applying the merge function in
Equation~\eqref{eq:mergeedges} to the conditional distributions:
\begin{dmath}
Pr((s_1,s_2).\boldsymbol{e}|s_1.\boldsymbol{l_1},s_2.\boldsymbol{l_2})\\ = {\left[m^{\{T,F\}}(\{p^{(r_1,r_2)}|r_i\in s_i\})\right]}((s_1,s_2).\boldsymbol{e} |s_1.\boldsymbol{l_1},s_2.\boldsymbol{l_2})
\end{dmath}
While node label and edge existence factors
are no longer over disjoint sets of variables, 
because  the 
label probabilities can be
computed before the edge probabilities, there are no
cyclic dependencies between random variables.  
Hence,
the product of these factors still is a normalized probability distribution, which allows us to use the new edge existence factors instead of the previous ones in the 
 full joint distribution \eqref{eq:pgm}, its factorization \eqref{eq:pgmpr}, and the marginal over all assignments that give rise to the same possible world graph  \eqref{eq:possworldprob}. 
Similar reasoning can be used to allow for more complex dependencies
between node labels and edges' existence, see Getoor et al.~\cite{getoor:jmlr02} for details.
While the probabilistic model can easily admit more complex correlations, efficient
indexing becomes more challenging, as we discuss in \secref{sec:algo-correlations}.


\section{Subgraph Pattern Matching}
\label{sec:subgraph-matching}
We now define the task of  subgraph pattern matching over uncertain graphs. 
Our discussion assumes undirected graphs, but our approaches are equally applicable to directed graphs. 
We start by defining a match of a query $Q$ in a graph $\G$ where there is no uncertainty, and we then define probabilistic subgraph pattern matching. A query graph $Q=(V_Q,E_Q)$ is a graph where each node $v\in V_Q$  is labeled with a label $l_Q(v)\in\Sigma$. 

\begin{mydef}
\textbf{Match:} Given a labeled graph $\G = (V_G, E_G, l_G(.))$ and a query graph $\Q = (V_Q,E_Q,l_Q(.))$, a subgraph $M = (V_M, E_M)$ of $\G$  is a match of $Q$ in $\G$ if and only if there is a bijective mapping $\psi: V_\Q \rightarrow V_\M$ such that (i) $\forall u\in V_\Q: l_Q(u)=l_G(\psi(u))$ and (ii) $(\psi(u),\psi(v)) \in E_\M$ if and only if $(u,v)\in E_\Q$.
\end{mydef}
\begin{mydef}
\textbf{Probabilistic Match:} Given a PEG $\T$ and a query graph $\Q$, a graph $M$ is a probabilistic match of $Q$ in $\T$   if and only if $M$ is a match of $Q$ in at least one legal possible world graph $\G$ of $\T$, that is, one where no two nodes share a reference. The probability of the match $M$ is the sum of the probabilities of all possible world graphs of $\T$ where $M$ is a match:
\begin{equation}
\label{eq:isomorphismprob}\displaystyle Pr(M) = \sum_{\G \in PW(\T) \land M \subseteq G} Pr(\G)
\end{equation}
\end{mydef}
\begin{mydef}
\textbf{Probabilistic Subgraph Pattern Matching:} Given a PEG $\T$, a query graph $\Q$, and a probability threshold $\alpha$, find all matches of $\Q$ in $\T$ whose probability $Pr(M)$ is greater than or equal to $\alpha$. 
\end{mydef}

Naively, this problem could be solved by performing subgraph pattern matching over each possible world graph and for each match found, summing the probabilities of possible worlds it appears in. Clearly, this approach is computationally infeasible. In the remainder of this section, we show how to (a) find all matches by performing subgraph matching on a single graph only, and (b) calculate the probability of a given match directly, without need to explicitly consider all possible worlds it appears in. This provides the basis for the algorithms  discussed in Section~\ref{sec:algorithms}, which further speed up probabilistic subgraph pattern matching. 

\vspace{2pt}
\noindent
\textbf{Finding Matches.} For a given PEG~$U$, let $\R$ be the graph that has a node for each $s\in S$, labeled with the set of labels~$L(s)$ that are associated with $s$ with non-zero probability, that is, $L(s) = \{l'| l' \in \Sigma \wedge Pr(s.l=l')>0\}$, and an edge between two nodes $s_1$ and $s_2$ if and only if $Pr((s_1,s_2).e=T)>0$. We generalize the notion of match to this case by requiring the query node label to be in the set of labels of the matched node. Clearly, if $M$ is a match in a legal possible world of $U$, it is a match in $G_U$. However, while all matches $M$ in $G_U$ are a match in some possible world of $U$, this world might not be legal. This is the case  if and only if the match includes two nodes that share a reference. We therefore further  extend the matching procedure on $G_U$ to not return matches where two nodes share a reference. This ensures that the matches on $G_U$ are exactly the probabilistic matches on $U$. For the discussions to follow, we use the term probabilistic entity graph to denote $G_U$ as well, as it is the structure that our algorithms operate on.

\vspace{5pt}
\noindent
\textbf{Calculating Probabilities.} In \eqnref{eq:isomorphismprob}, the probability of a match~$M$ found on $G_U$ is defined based on a set of possible world graphs, summing their probabilities as given by \eqnref{eq:possworldprob}. The graphs in this set  are exactly  those containing all nodes in $V_M$ with correct labels as well as all edges in $E_M$, and arbitrary sets of additional nodes and edges. Thus, the probability of $M$ can be rewritten as the marginal 
\begin{align}
Pr(M) &= Pr_n(M) \cdot Pr_{le}(M)\label{eq:pmatch}\\
Pr_n(M) &= Pr(V_M.n=T)\\
Pr_{le}(M) &= \prod_{v\in V_M} Pr(v.l=l(v)) \cdot \prod_{e\in E_M} Pr(e.e=T)
\end{align}
where  $Pr(V_M.n=T)$ is the corresponding marginal of $Pr(S.n)$ that sums out values of all node existence variables whose nodes are not part of~$M$. 
In practice, as in  \eqnref{eq:splitnodeprob}, we exploit independencies in the underlying graphical model to calculate this probability
as a product of existence probabilities of smaller sets of nodes. Recall that $\mathcal{C}(S.n)$ partitions the set of node existence random variables $S.n$ based on the connected components of the Markov network. As each node in a match corresponds to one such random variable, we can use the same partitioning, restricted to the set of nodes $V_M$ in the match, to calculate $Pr(V_M.n=T)$ as 
 $\prod_{C.n\in\mathcal{C}(S.n) }Pr((V_M.n\cap C.n)=T)$.  
Note that $Pr_{le}(M)$ is \emph{subgraph decomposable}, that is, for two disjoint subgraphs $M_1$ and $M_2$, $Pr_{le}(M_1)\times Pr_{le}(M_2)=Pr_{le}(M_1 \cup M_2)$, while this is not the case for $Pr_n(M)$.

\section{Algorithms}
\label{sec:algorithms}
The problem of probabilistic subgraph pattern matching with identity uncertainty is \#P-complete. To increase efficiency, we propose a new  \emph{path-based} approach to find probabilistic matches of queries. Our approach decomposes the query into a set of paths, finds matches of individual paths, and exploits probabilistic information to prune the space of possible matches.

By focusing on paths rather than nodes when finding candidate matches,
we can better exploit probabilistic information for pruning. If we
considered only the probabilities associated with the nodes as the
criteria for candidacy (as opposed to paths), the search space would
end up being very large, because node probabilities tend to be much
larger than the final query probability, leading to a search space with many more false positives.
On the other hand, a path-based approach has better pruning capabilities, especially when used in association with \emph{path context information} and further reduction techniques as outlined in the following paragraphs.

\begin{figure*}
\begin{center}
\includegraphics[scale=0.4]{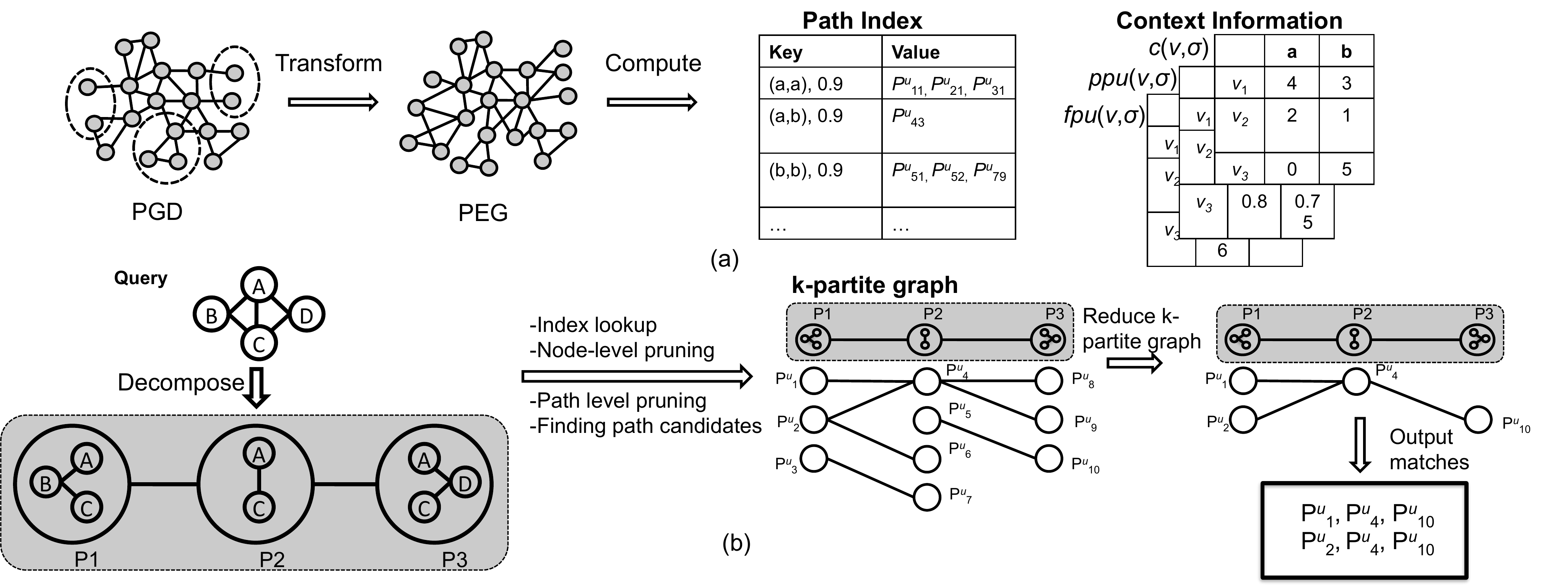}
\end{center}
\vspace{-10pt}
\caption{(a) Offline and (b) online phase schematic diagrams} 
\vspace{-10pt}
\label{fig:schematic}
\end{figure*}

In order to enable efficient and scalable online processing, we divide
the work of answering probabilistic subgraph pattern matching queries
into an offline and an online phase, which are summarized in Figures
\ref{fig:schematic}(a) and \ref{fig:schematic}(b), respectively. 
The offline phase first precomputes entity-level probability information. 
Second, it builds a novel disk-based \emph{context-aware path index} on the probabilistic entity graph, indexing not only all the paths in the PEG up to a given length, but also other context information that captures different properties of the path local neighborhoods (\secref{sec:offline}). The online phase answers the online user's query (\secref{sec:query-processing}). It first decomposes the query into paths and then constructs a search space over the paths in three steps, by 1) accessing the path index to find an initial set of path \emph{candidates}, i.e., paths in the PEG that can potentially be a match for the paths in the decomposition, 2) employing context information to prune the sets of candidates for all query paths, and 3) reducing the search space for full graph matches  
using a technique called \emph{reduction by join-candidates}, which performs message passing in a k-partite graph where each partition corresponds to a path in the query decomposition. This results in the final search space, from which the algorithm then generates the actual matches. \forlongversion{Notations used in this Section are list in \tabref{tab:notations2}.}

\forlongversion{
\begin{table}
\begin{tabular}{|p{0.8in}| p{2.25in}|}
\hline
Notation & Definition\\
\hline
$P$ & Query path\\
$P^u$ & Entity graph path\\
$\psi(v)$ & Entity graph node matching the query node $v$ \\
$l(v)$ & Label of node $v$\\
$V_M$ & Set of nodes of subgraph $M$\\
$E_M$ & Set of edges of subgraph $M$\\
$Pr_{le}(M)$ & Probability of the node label and edge existence components of subgraph $M$\\
$Pr_{n}(M)$ & Probability of the entity node existence component of subgraph $M$\\
$\alpha$ & Input query threshold\\
$\beta$ & Path index construction threshold\\
$\gamma$ & Path index resolution coefficient\\
$L$ & Path index maximum path length\\
$\Gamma(v)$ & Neighbors of node $v$\\
$\refs(v)$ & Set of underlying references of node $v$\\
$c(v,\sigma)$ & Cardinality of node $v$ with respect to $\sigma$\\
$ppu(v,\sigma)$ & Partial probability upperbound of node $v$ with respect to $\sigma$\\
$fpu(v,\sigma)$ & Full probability upperbound of node $v$ with respect to $\sigma$\\
$\mathcal{P}$ & Set of paths in a path decomposition\\
$JP(P_1,P_2)$ & Join predicates between $P_1$ and $P_2$\\
$J(P_1)$ & Paths that join with $P_1$\\
$cn(P)$ & Set of candidates of path $P$\\
$cn(P_1,P_1^u,P_2)$ & Set of paths in $cn(P_2)$ that are candidates to be joined with $P_1^u \in cn(P_1)$\\
\hline
\end{tabular}
\caption{Notations used in \secref{sec:algorithms}}
\label{tab:notations2}
\end{table}
}
\subsection{Offline Phase}
\label{sec:offline}
\coloredcomment{\color{blue}Should we explicitly mention that $G_U$ is built as part of the offline phase, or is it given from Section 4? If we mention it here again, maybe we need to give it a name.}

The offline phase precomputes the following pieces of information over the probabilistic entity graph: component probabilities, path index, and context information on nodes. A schematic diagram of the offline phase steps is shown in \figref{fig:schematic}(a). We discuss each type in the following subsections.

\vspace{5pt}
\topic{Component Probabilities:}
Calculating match probabilities requires evaluating ~\eqnref{eq:pmatch}. To reduce calls to the PGM engine during online inference, we precompute and store the underlying probabilities. As $Pr_{le}(.)$ is decomposable, we only precompute its parts, that is, node label and edge existence probabilities, by applying the corresponding merge functions on the underlying input distributions as specified in Equations \ref{eq:mergelabels} and \ref{eq:mergeedges}, respectively. Since $Pr_n(.)$ is not decomposable, we precompute node existence marginals for all possible valid configurations of every connected component, i.e., those consisting of entities not sharing a reference. In general the connected components are expected to be small enough in practice for this to be feasible. If not, we could instead either employ an approximate inference technique to compute the marginals, or compute them on demand using the PGM engine.

\vspace{5pt}\topic{Path Index:}
The path index contains all paths in the probabilistic entity graph that have length at most~$L$, probability at least~$\beta$\footnote{Paths with smaller probability are computed on demand.}, and do not contain two nodes sharing an underlying reference. 
Every entry in the path index consists of the following information:
\begin{myitemize}
\item \textbf{Key:} the entry's key is a pair $\langle \textbf{X}, \pi \rangle$, where $\textbf{X}\in\Sigma^{l+1}$ is a sequence of node labels of length $l+1$, and $\pi\in \{\beta, \beta+\gamma, \beta+2\gamma,\ldots,1\}$ is a probability value. The parameter
$\gamma$ defines the resolution of the index and provides a tradeoff between the accuracy and the query response times of the index. 
\item \textbf{Value:} the entry's value is the set of paths $\mathcal{P}^u$ of length $l$ with probability under the node label assignment $\textbf{X}$ between $\pi$ and $\pi+\gamma$ satisfying the reference constraint. For every $P^u \in \mathcal{P}^u$, we store the path itself as well as its two probability components $Pr_{le}(P^u)$ and $Pr_{n}(P^u)$. 

\end{myitemize}


\forlongversion{\vspace{5pt}\noindent\textbf{Example:} If a path $P^u=(1,2,3)$ has the probabilities $Pr_{le}(P^u)=0.9$ and $Pr_{n}(P^u)=1.0$ under the node label assignment $(x_1, x_2, x_3)$, then assuming an index resolution $\gamma=0.1$, this path will be associated with the key $\langle (x_1, x_2, x_3), 0.9 \rangle$. }

To increase efficiency, we build a \emph{two-level index}, where the first level, accessing $\textbf{X}$ via equality predicates, is a hash index, and the second level, accessing $\pi$ via range predicates, is a B+ tree index.
Index construction starts with paths consisting of a single node ($l=0$) and builds entries of length $l+1$ based on those of length $l$, exploiting the fact that all paths with probability at least $\beta$ must consist of sub-paths with probability at least $\beta$ as well. 
We exploit the fact that entries for different label sequences of the same length can be constructed independently to build those \emph{in parallel} using multiple threads. We use a synchronization barrier to ensure that all  paths of the current length have been indexed before proceeding to the next length.  To increase I/O performance, we accumulate a group of records in a \emph{memory buffer} before writing the buffer to disk. Finally, for undirected graphs, entries for labels $\textbf{X}=\{X_1,X_2,\ldots,X_{l-1},X_l\}$ are identical to those for $\textbf{X'}=\{X_l,X_{l-1},\ldots,X_2,X_1\}$ because of \emph{symmetry}, and we therefore  only store one direction for each such case and derive the other one as needed.

\vspace{5pt}\topic{Context Information:}
\label{sec:context}
When pruning the set of candidate matches for a path in \secref{finding-path-candidates}, we rely on context information for nodes, which is the third type of information precomputed during the offline phase. For a node $v \in \R$ and a label $\sigma\in \Sigma$,
     let $N(v,\sigma)$ be the set of neighbors of $v$ that have $\sigma$ in their set of possible labels, i.e., \\[2pt]
\centerline{$N(v,\sigma)=\{v'|{v'\in \Gamma(v),\sigma \in L(v), \refs(v)\cap \refs(v')=\emptyset}\},$} \\[2pt]
where $\Gamma(v)$ is the set of neighbors of node $v$, and $\refs(v)$ is the set of underlying references of node $v$.
For each node $v \in V_\T$ and  label $\sigma\in \Sigma$, we compute the following values:
\begin{myitemize}
\item \textbf{Cardinality} $c(v,\sigma)$, which is simply the size of $N(v,\sigma)$:\\[2pt] \centerline{$c(v,\sigma)=|N(v,\sigma)|$}
\item \textbf{Partial Probability Upperbound} $ppu(v,\sigma)$, which is an upperbound for the probabilities in the neighborhood of $v$ considering only the edges between $v$ and $N(v,\sigma)$. \\[2pt] \centerline{$ppu(v,\sigma)=max_{v' \in
    N(v,\sigma)}Pr((v,v').e=\bT)$}
\item \textbf{Full Probability Upperbound} $fpu(v,\sigma)$, which is an upperbound for the probabilities in the neighborhood of $v$ 
also taking into account the neighbors' labels. \\[2pt]\centerline{$fpu(v,\sigma)=max_{v' \in N(v,\sigma)}Pr(v'.l=\sigma)\cdot Pr((v,v').e=\bT)$}
\end{myitemize}
These measures capture different aspects of node/path neighborhoods. 
During the online phase (cf.~\secref{finding-path-candidates}), we use a combination of the cardinality and full probability upperbound to prune path candidates at the individual node level, and a combination of full and partial probability upperbounds to prune path candidates at the entire path level.

\forlongversion{
\begin{figure}
\begin{center}
\includegraphics[scale=0.3]{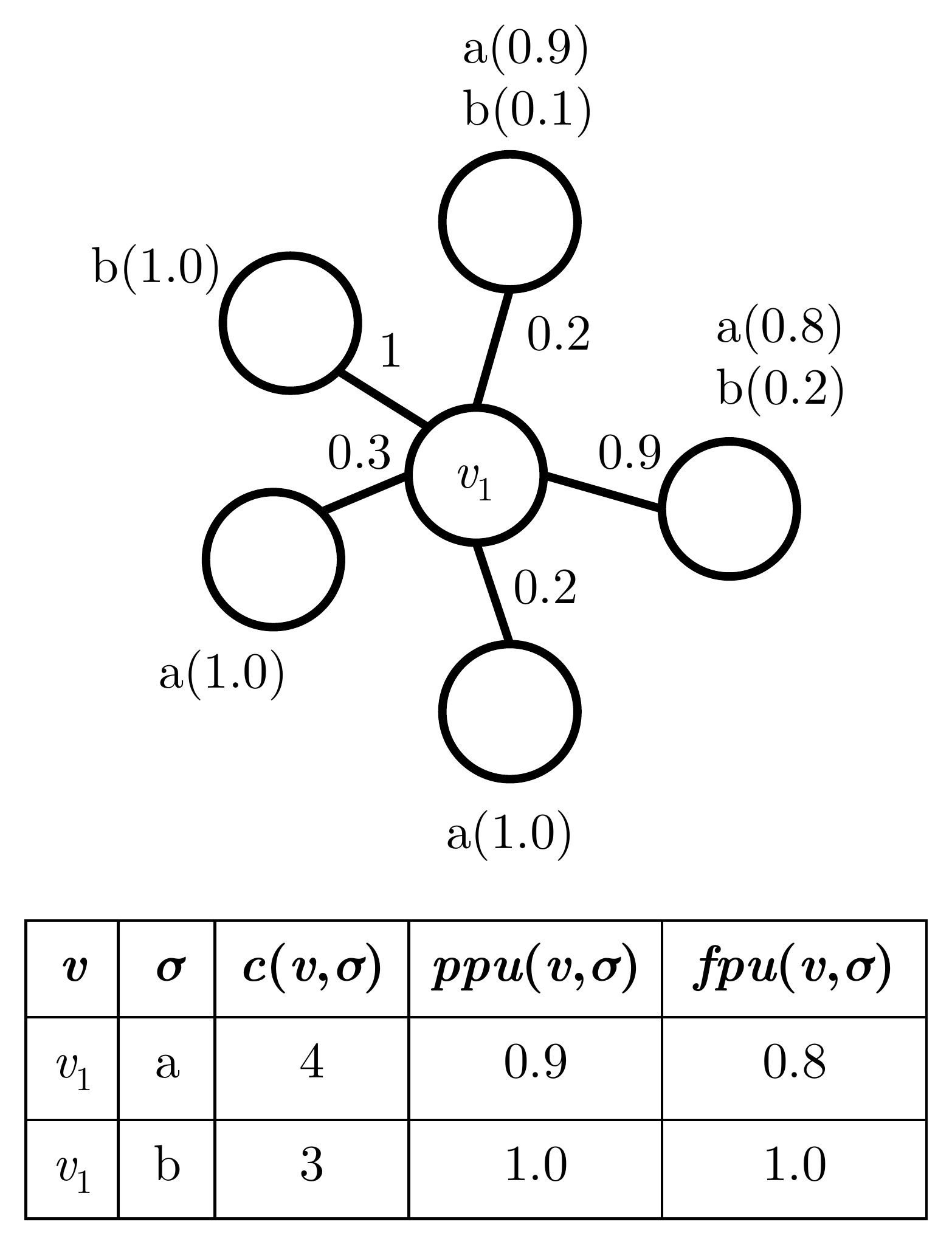}
\end{center}
\caption{Context information example}
\label{fig:context-information-example}
\end{figure}

\vspace{5pt}\noindent\textbf{Example:} In \figref{fig:context-information-example}, $c(v_1,a)=4, c(v_1,b)=3$. $ppu(v_1,a)=0.9$ because the highest edge probability that connects $v_1$ to a node with label $a$ is $0.9$. Similarly, $ppu(v_1,b)=1.0$. Finally, $fpu(v_1,a)=0.72$ because it has an edge with existence probability of $0.9$ connecting it to a node with probability of $0.8$ for the label $a$. Similarly, $fpu(v_1,b)=1.0$.
}

\subsection{Online Phase}
\label{sec:query-processing}
Our online query processing technique consists of five main steps: decomposing the query into a set of paths, obtaining a set of candidates for every path in the decomposition, obtaining \emph{join-candidate paths} for every candidate path, which are candidate paths whose query paths share a node with the given candidate and can thus extend it to form a partial match, jointly reducing the candidate search space by \emph{reduction by join-candidates}, and finally finding matches to the full query. A schematic diagram of the online phase steps is shown in \figref{fig:schematic}(b). Below we discuss each step in detail.
\subsubsection{Path Decomposition}
The task of \emph{path decomposition} is to split the query into a set of possibly overlapping paths, each of length $L$ or less, that cover the entire query, and whose matches can be obtained from the path index. To preserve the structural information of the query, intersection points between the paths are expressed as join predicates, which have to be satisfied when combining path matches into a full query match. For example, \figref{fig:schematic}(b) shows a query and its decomposition into three paths $P_1$, $P_2$ and $P_3$. In order to preserve the structural information of $Q$, any three paths $(P_1^u, P_2^u,P_3^u)$ that match $(P_1, P_2,P_3)$ must satisfy the predicates $P_1^u.A=P_2^u.A$, $P_1^u.C=P_2^u.C$, $P_3^u.A=P_2^u.A$, and$P_3^u.C=P_2^u.C$ (we use $P_1^u.A$ to denote the vertex in path $P_1^u$ that matches the vertex $A$ in path $P_1$ in the query). Query path decomposition thus decomposes a query $Q$ into a set of node/edge overlapping paths $\mathcal{P}$. For every pair of overlapping paths $P_1$ and $P_2$, the decomposition defines a set of \emph{join predicates} $JP(P_1,P_2)$. Further, we denote the set of paths joining with a path $P$ by $J(P)$.

Since a single query has multiple valid path decompositions, and each decomposition may lead to a different query processing cost, we would like to find a least-cost path decomposition. Ideally, the cost of a decomposition should express the number of operations involved in order to produce the final query results. 
As the intricacy of the algorithm makes it difficult to calculate such a number, 
 we instead use an estimate of  the initial query search space size $SS_0$. We would thus like to find $argmin_{\mathcal{P} \subseteq \mathbb{P}(Q), \mathcal{P} \text{ covers } Q} SS_0(\mathcal{P})$, where $\mathbb{P}(Q)$ is the set of all possible paths of length at most $L$ in $Q$. More specifically, for each path $P$ in the decomposition, we estimate the number of matches,  or its cardinality $C(P,\alpha)$, as discussed below. We then estimate the search space size as the product of all such path cardinalities. 
The cardinality is based on  the number of database paths matching the query path $P$ with probability at least $\alpha$, but also takes into account the fact that those matches will have to be extended to neighboring query paths. We therefore express $C(P,\alpha)$ in terms of the following quantities.
\begin{myenumerate}
\item \textbf{Number of candidates} $|PIndex(l_Q(V_P), \alpha)|$ 
matching $P$'s label sequence $l_Q(V_P)$ with probability at least $\alpha$ in the path index. 
\item \textbf{Path degree} $degree(P)$: sum of path node degrees, not counting edges on the path, that is,  \\[2pt]\centerline{$degree(P)=\sum_{n \in V_P}degree(n)-2\times length(P)$}
\item \textbf{Path density} $density(P)$: this measures 
how close the nodes on $P$ are to forming a clique. Let $K$ be the number of edges between the nodes of $P$, and $M$ 
the number of nodes on the path, then $density(P)= \frac{2K}{M(M-1)}$.
\end{myenumerate}

\forlongversion{
\begin{figure}
\begin{center}
\includegraphics[scale=0.26]{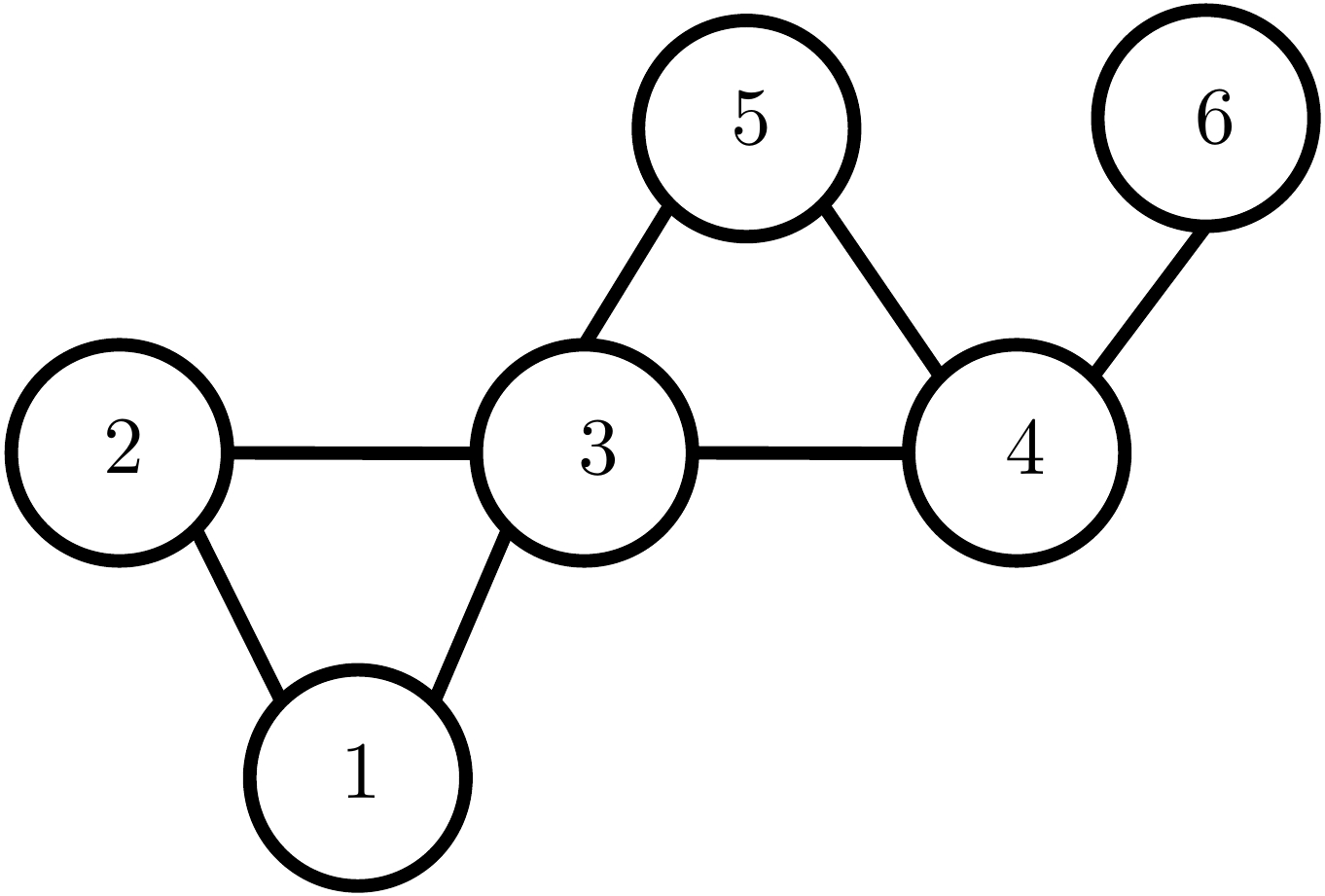}
\end{center}
\caption{Path degree and density example}
\label{fig:path-degree-density}
\end{figure}
\vspace{5pt}\noindent\textbf{Example: } The path degree of path $(1,2,3,4)$ shown in \figref{fig:path-degree-density} is 5, and its density is $4/6$.
}

Taking into account the direction of influence of these components on the true number of matches, we approximate $|P|$ as:
$$C(P,\alpha) \propto \frac{|PIndex(l_Q(V_P), \alpha)|}{degree(P)\cdot density(P)}$$
Therefore, our goal is to find 
$$argmin_{\substack{\mathcal{P} \subseteq \mathbb{P}(Q),\\ \mathcal{P} \text{ covers } Q}} \prod_{P\in\mathcal{P}}\frac{|PIndex(l_Q(V_P), \alpha)|}{degree(P)\cdot density(P)}$$
Since it is not practical to query the index for an arbitrary $\alpha$ and $l_Q(V_P)$ at query time, we build a histogram for every possible label sequence $\textbf{X}$ during the offline phase at selected probability points $(\alpha_0,\ldots,1)$. 
At runtime, we use exponential curve fitting to estimate the value of $|PIndex(l_Q(V_P), \alpha)|$ given $hist(l_Q(V_P), \alpha_i)$ and $hist(l_Q(V_P), \alpha_{i+1})$ where $\alpha_i<\alpha<\alpha_{i+1}$.

We reduce the problem of optimizing the cost function to that of SET COVER, 
where the set of query edges corresponds to the universal set (in the corresponding SET COVER instance), and each path $P$ in the query with length at most $L$ is a candidate set. Note that we allow  paths with shared edges, as this can reduce the cost of several paths at once (e.g., in the case of a very selective edge connected to multiple non-selective paths).
The cost of the cover is the product of the individual costs of the participating paths. 
Since SET COVER is NP-complete, we use the standard greedy approximation to solve the problem, which calculates an \emph{efficiency} metric for every path by dividing its length by its cost, and then greedily adds the path with the highest efficiency to the cover, continuing iteratively until all the query edges are covered.
 
\subsubsection{Finding Path Candidates}
\label{finding-path-candidates}
Given a path decomposition $\mathcal{P}$, the next step is to find candidate matches for every query path. Therefore, at a high level, for every path $P\in \mathcal{P}$, we access the path index to get its matches $PIndex(l_Q(V_P), \alpha)$, by only keeping those paths that satisfy certain context criteria. We denote the resulting set of matches by $cn(P)$ ($\subseteq PIndex(l_Q(V_P), \alpha)$). This second step relies on the following query statistics: 

\begin{myitemize}
\item \textbf{Node-level statistics:} For every node $n\in V_Q$, we calculate its neighborhood label count for every label $\sigma \in \Sigma$, \\[2pt]\centerline{$c(n, \sigma)=|\{m|{m\in \Gamma(n), l_Q(m)=\sigma}\}|$}

\item \textbf{Path-level statistics:} For every path $P \in \mathcal{P}$, we collect information on its neighboring nodes in the query, the nodes on $P$ these neighbors are connected to, and the query edges outside $P$ that connect nodes on $P$. In order for a path match to be a candidate for contributing to a full query match, it has to be possible to extend this match to at least this neighborhood, and we can safely prune other path matches. More specifically, we use the following information:
\begin{myenumerate}
\item Path neighbors $\Gamma(P)$: the set of nodes that are not on $P$ but are neighbors of at least one node on $P$.
\item Reverse path neighbors: for every $m\in \Gamma(P)$, $rv(P,m)$ is the set of nodes on $P$ that are neighbors of $m$.
\item Path cycles: for every $n\in V_P$, path cycles, $cyc(P,n)$, is the set of nodes on $P$ that are also connected to $n$ by a query edge outside the path, and thus appear together with $n$ in a cycle. To avoid information duplication, each such edge only contributes to the path cycles of one of its endpoints.


\end{myenumerate}
\forlongversion{\vspace{5pt}\noindent\textbf{Example: } In \figref{fig:path-degree-density}, path neighbors of the path $(1,2,3,4)$ are the set of nodes $\{5,6\}$. Reverse path neighbors of node $5$ are $\{3,4\}$. There is a path cycle formed by the edge between the nodes $1$ and $3$.}
\end{myitemize}
\textbf{Node-level pruning:}
Using the node-level statistics, we calculate a set of candidates $cn(n)$ for every node $n\in V_Q$ as follow: 
\begin{myenumerate}
\item For every label $\sigma\in\Sigma$, $v$ must have a number of neighbors that is greater than or equal to the number of neighbors of $n$ with label $\sigma$, i.e., $c(v,\sigma) \ge c(n,\sigma), \forall \sigma \in\Sigma$.
\item For every label $\sigma \in\Sigma$, the probability of $v$ having the correct label and at least the number of neighbors labeled $\sigma$ required by the query has to exceed the query threshold $\alpha$. Using precomputed full probability upperbounds as approximation and taking into account multiple occurrences of the same label, we therefore further restrict candidates $v$ for $n$ to those satisfying $Pr(v.l=l_Q(n))\times fpu(v,\sigma)^{c(v,\sigma)} \ge \alpha, \forall \sigma \in\Sigma$. 
\end{myenumerate}
\textbf{Path-level pruning:}
Next, we prune the set of candidate paths using path-level statistics. 
For each path $P^u \in PIndex(l_Q(V_P), \alpha)$, we perform the following tests:
\begin{myenumerate}
\item For every node $v \in V_{P^u}$, $v$ must be a candidate for the corresponding node $n$ in $P$, i.e., $v \in cn(n)$.
\item The probability of a path together with its neighboring nodes and cycles 
must be greater than or equal to $\alpha$, which we test using $(Pr_{le}(P^u)\times Pr_{n}(P^u))\times pu(P^u) \times cpr(P^u) \ge \alpha$, with $pu(P^u)$ and $cpr(P^u)$ defined as follows.

The \emph{path-neighborhood probability upperbound} $pu(P^u)$ of a candidate path $P^u$ matching a query path $P$ is an upperbound for the probability of all nodes matching $\Gamma(P)$ and their edges. 
Let $m \in \Gamma(P)$ be a path $P$ neighbor, and $n$ a node on $P$ such that $n \in rv(P,m)$. We compute a probability upperbound $pu(n,m,P^u)$ 
on the neighborhood of $m$ as:
\begin{equation*}
fpu(\psi(n),l_Q(m))\prod_{n'\in rv(P,m), n' \neq n}ppu(\psi(n'),l_Q(m))
\end{equation*}
where we use the full probability upperbound $fpu$ for the edge between the match of $m$ and the selected neighbor $n$, and partial probability upperbounds for all other neighbors of $m$'s match, thus ensuring that information on $m$ is only considered once.
Choosing the tighest upperbound over all reverse path neighbors $rv(P,m)$ and aggregating 
over all $m \in \Gamma(P)$, we get the overall path $P^u$ neighborhood probability upperbound: 
\\[2pt]\centerline{$pu(P^u)=\prod_{m\in \Gamma(P)} min_{n \in rv(P,m)}pu(n,m,P^u)$}



The \emph{path-cycles probability} $cpr(P^u)$ is the overall probability of edges not on the path $P^u$ but connecting path nodes:
\\[2pt]\centerline{$cpr(P^u)=\prod_{\substack{n\in V_P,\\ m\in cyc(P,n)}}Pr((\psi(n),\psi(m)).e = T)$}
\end{myenumerate}

Finally, for every path $P$ in the decomposition,  we obtain the list of candidates $cn(P)$ that contains exactly those paths from the initial set $PIndex(l_Q(V_P), \alpha)$ 
 that pass the above tests. 

\subsubsection{Finding Join-Candidates}
In this step, for every candidate path $P^u \in cn(P)$ of every query path $P$, we find a set of paths that are candidates to be joined with $P^u$.
Recall that every query path $P_1\in \mathcal{P}$ can be joined with a set of paths $J(P_1) \subseteq \mathcal{P}$, and there is a set of join predicates $JP(P_1,P_2)$ between $P_1$ and every path $P_2 \in J(P_1)$. For a query path $P_1\in \mathcal{P}$, and a candidate path $P_1^u\in cn(P_1)$, we define its join-candidate paths of type $P_2\in J(P_1)$ as:
\begin{dmath*}
cn(P_1,P_1^u,P_2)={\{P_2^u|P_2^u \in cn(P_2)} {\land jp(P_1^u,P_2^u)=\bT, \forall jp \in JP(P_1,P_2)} \\ {\land Pr(P_1^u \circ P_2^u)\ge \alpha\ \land \refs(V_{P_1^u})\cap \refs(V_{P_2^u})=\emptyset}\}
\end{dmath*}
where $jp(P_1^u,P_2^u)$ is the instantiation of the predicate $jp\in JP(P_1,P_2)$ using paths $P_1^u$ and $P_2^u$, and $P_1^u \circ P_2^u$ is 
the subgraph consisting of the two joined paths. Intuitively, $cn(P_1,P_1^u,P_2)$ refers to the set of paths in $cn(P_2)$ that are candidates to be joined with $P_1^u \in cn(P_1)$.

To facilitate finding join-candidate paths, for each 
$P\in \mathcal{P}$, while finding $cn(P)$, we build a  lookup table $T(P,P_i)$ for each query path $P_i\in J(P)$. For every table $T(P,P_i)$,  the  set of positions $\langle p_{i1},\ldots,p_{ik}\rangle $ indicates the nodes in $P_i$ that participate in join predicates.  The key for table $T(P,P_i)$ is a set of nodes $\langle n_1,\ldots,n_k \rangle$, and the values are paths in $cn(P)$ that have nodes $\langle n_1,\ldots,n_k \rangle$ at positions $\langle p_{i1},\ldots,p_{ik}\rangle $. Given a path $P_i^u\in cn(P_i)$, paths in $P$ which are joinable with $P_i^u$ can now be obtained using a direct lookup operation from table $T(P,P_i)$, where the access key is obtained from $P_i^u$.

\subsubsection{Joint Search Space Reduction}
\label{sec:join-reduction}
Joint search space reduction exploits the mutual relationship between the candidates and their join-candidates to reduce the size of all candidate lists before constructing full query matches, based on the following two observations. First, for a candidate match
of a path $P$ to contribute to a full query match, we must be able to combine it with at least one candidate for all query paths joining $P$. Second, if we can obtain an upperbound on the probability of all full query matches a candidate path can appear in, we can prune candidate paths based on the  query threshold $\alpha$. We refer to these two principles as \emph{reduction by structure} and \emph{reduction by upperbounds}, respectively, and discuss their details below. As they influence each other, the overall algorithm for joint search space reduction iterates between them until no further changes occur. 

We implement the reduction algorithm based on a \emph{k-partite graph}, where each partition corresponds to a query path, each vertex to  a candidate path match, and each link to a join between two candidate paths.\footnote{To avoid confusion, we use the terms (vertex/link) when referring to the k-partite graph, and (node/edge) when referring to the PEG.} Pruning a candidate thus corresponds to deleting a vertex and its outgoing links from the k-partite graph.

\begin{mydef}
\textbf{Candidate k-partite Graph}
A candidate k-partite graph is a k-partite graph that has a partition for each $P \in \mathcal{P}$, where the set of vertices of each partition $P$ are $cn(P)$. There is a link between $P_1^u$ in partition $P_1$ and $P_2^u$ in partition $P_2$ iff $P_2^u \in cn(P_1, P_1^u, P_2)$ (of course, $P_2^u \in cn(P_1, P_1^u, P_2) \iff P_1^u \in cn(P_2, P_2^u, P_1)$).
\end{mydef}

Every match of the query in the PEG corresponds to a subgraph of the candidate k-partite graph with one vertex per partition (i.e., one match for each query path) and all join links between them. We can thus safely prune all vertices that have no links to a partition they should link to, as well as those that cannot participate in any match with probability above the query threshold. 

\vspace{5pt}
\noindent \textbf{Reduction by structure.} Reduction by structure
removes vertices from the candidate k-partite graph by 
repeating the following step until no further changes take place: 
If a vertex has no links to at least one partition its query path
joins with, remove the vertex and all of its links to vertices in all
partitions.

\vspace{5pt}
\noindent\textbf{Reduction by upperbounds.}
In order to exploit probabilistic information during search space reduction, we now introduce two types of vertex weights, based on $Pr_{le}(.)$ and $Pr_n(.)$, respectively, and then discuss a message passing scheme that exploits these weights to obtain bounds for reduction by upperbounds. 

The first type of weights is assigned 
such that when a subgraph's weights are multiplied, we obtain the final $Pr_{le}(.)$ probability of the corresponding match. 
To avoid double contributions in cases of overlap between paths, 
we assign the overlapping elements' probability to exactly one partition, i.e., for every $v \in V_\Q, e\in E_\Q$, we choose exactly one partition to cover $v$'s or $e$'s probability.  That is, if $v$ or $e$ exclusively belongs to one query path, it is assigned to the partition representing that path, and if $v$ or $e$ appear on multiple query paths, only one of their partitions is picked. Let partition $P$ (we use $P$ to refer to both the path and its corresponding partition) exclusively cover nodes and edges $cv(P)$ and $ce(P)$, respectively, then a vertex's first weight is 
$${w}_1(P^u)=\prod_{n \in cv(P)} Pr(\psi(n).l=l_Q(n)) \prod_{e \in ce(P)} Pr(\psi(e).e=\bT)$$
\noindent where $\psi(n)$ is the PEG node matching the query node $n$. As identity probabilities $Pr_n(P^u)$ are not decomposable, we directly use the identity probability of a path as the second weight of its corresponding vertex in the k-partite graph
(however, we cannot multiply weights of this type together as it is the case with $w_1$ weights):\\[2pt]
\centerline{$w_2(P^u)=Pr_n(P^u)$}

In addition to the two weights, each vertex $P^u$ has an associated \emph{perception vector} of length $k$, that is, with one entry per partition. Each entry is an upperbound on the $w_1$ weights of all vertices in that partition that can appear in a full match with $P^u$. Initially, we have $w_1(P^u)$ for the entry corresponding to $P^u$'s own partition, and $1$ for all other partitions. During message passing, each vertex first sends its current vector to each of its neighboring vertices (excluding the entry for the receiving neighbor's partition). Once all messages are received, each vertex $P_1^u$ updates its own vector based on the values received from its neighbors as follows. For each vector entry corresponding to a partition $P$ and each partition $P_2$ containing neighboring nodes of $P_1^u$, we choose the maximum value for $P$ sent by the neighbors in $P_2$. We then take the minimum of these over all such $P_2$ as the new value in the vector, and iterate the overall process. The upperbound used to prune a  vertex (and thus a  candidate path) based on the query threshold $\alpha$ then is the product of all entries in the vertex' vector  and its weight $w_2$.

As discussed above, the final algorithm iterates between both types of reduction until no further changes take place. We further improve efficiency by avoiding unnecessary updates and exploiting parallelism, as discussed next.

\vspace{5pt}
\noindent \textbf{Incremental maintenance.} 
We only recompute upperbounds for vertices for which a neighbor has been deleted or has reduced its perception, and
only consider vertices connected to a newly deleted link for deletion.

\vspace{5pt}
\noindent \textbf{Parallel Implementation.} We develop a shared-memory parallel implementation for the reduction algorithm, with one thread per partition. 
We introduce appropriate locking protocols to avoid incorrect modifications of the k-partite graph by multiple threads at the same time. We note here that in addition to the parallel implementation of the reduction algorithm, we also exploit parallelism in other parts of the system such as constructing node candidates, path candidates, and building join-candidate sets.
 
\forlongversion{
\begin{figure*}
\begin{center}
\begin{tabular}{c c}
\includegraphics[scale=0.25]{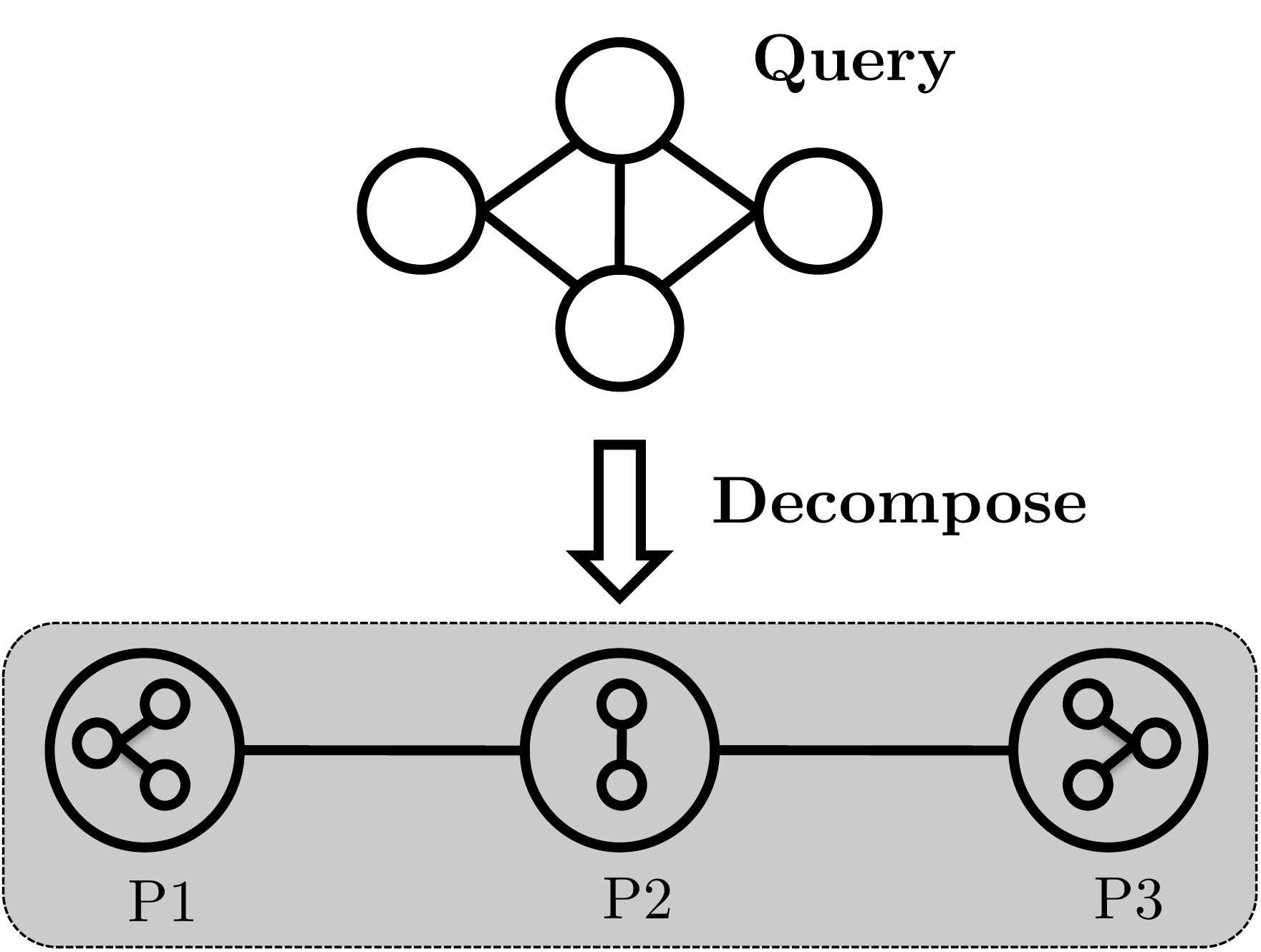}
&
\includegraphics[scale=0.2]{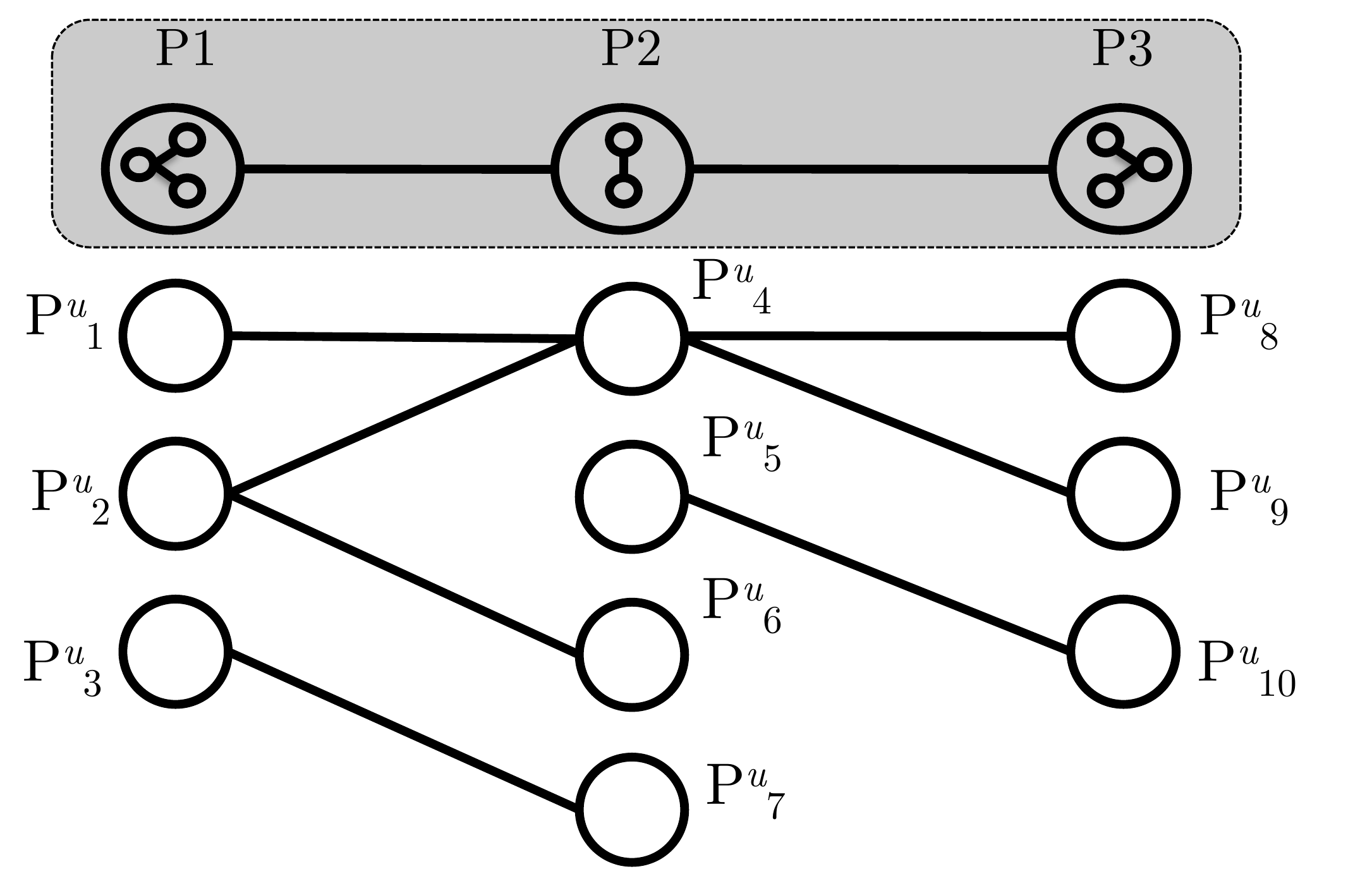}
\\
(a)
&
(b)
\\
\includegraphics[scale=0.2]{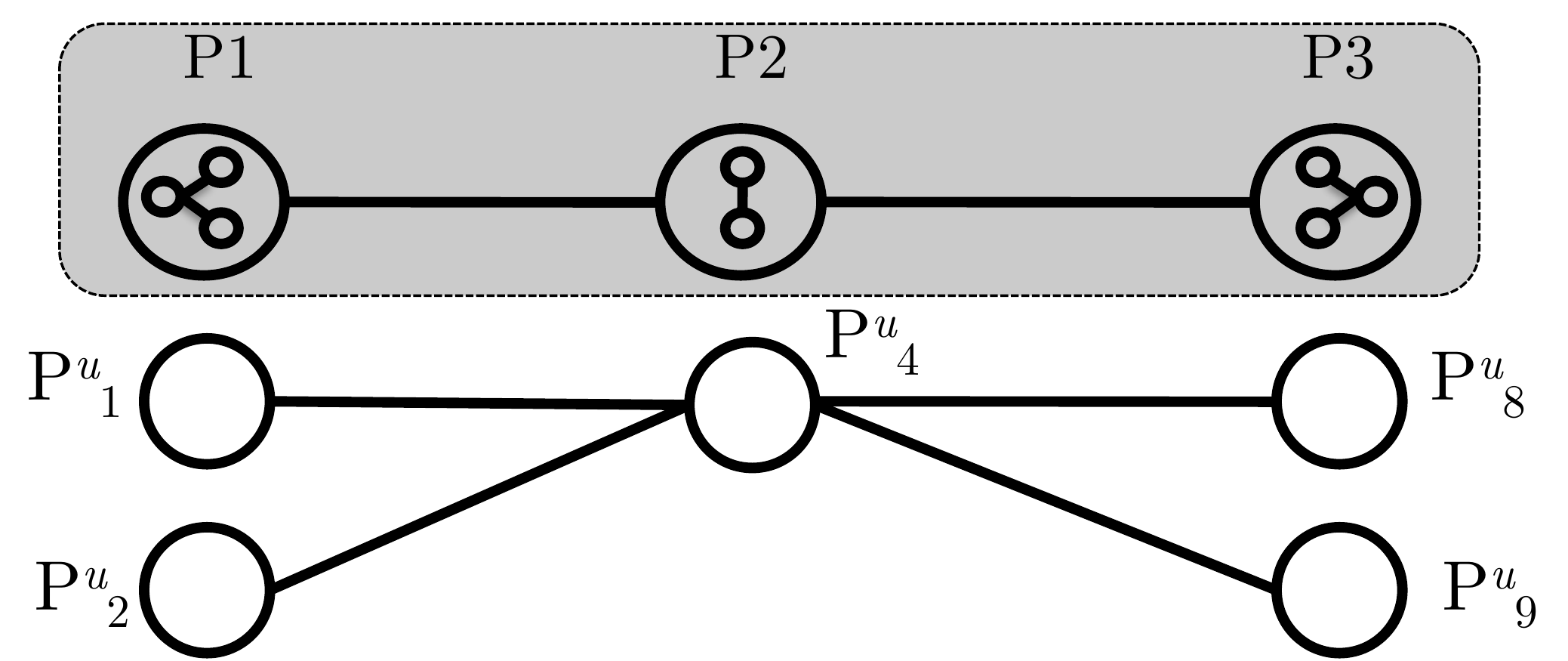}
&
\includegraphics[scale=0.2]{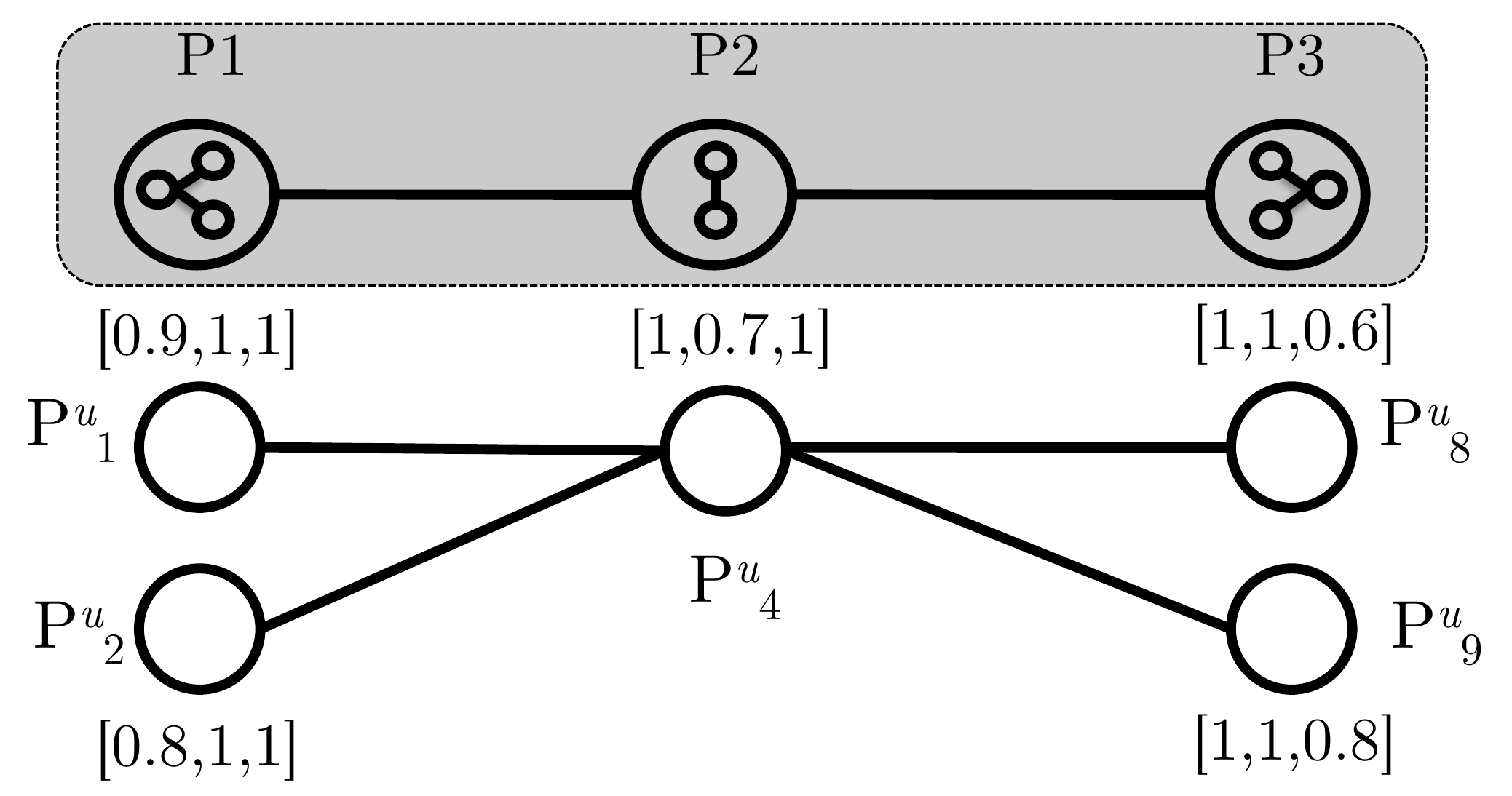}
\\
(c)
&
(d)
\\
\includegraphics[scale=0.2]{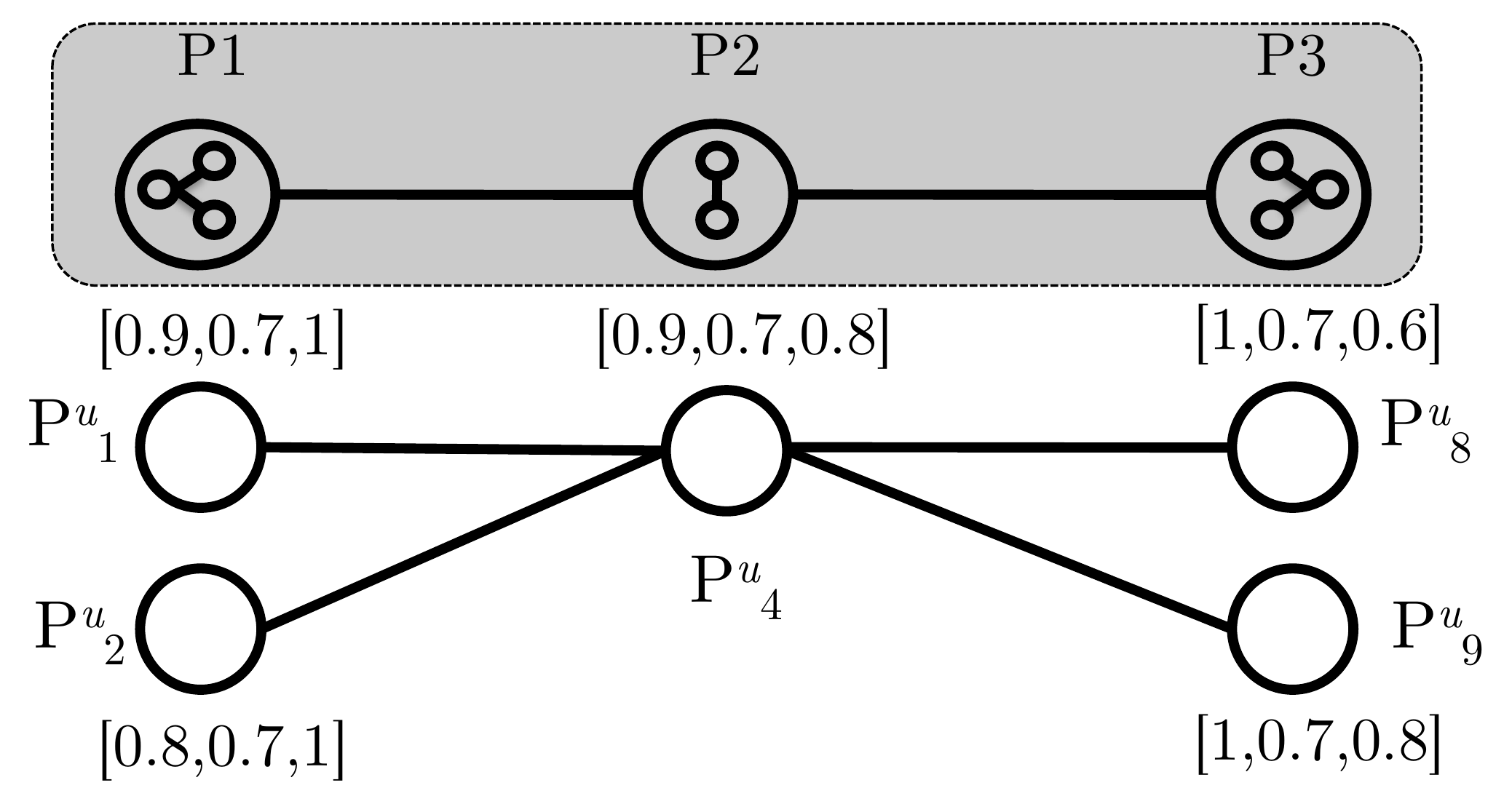}
&
\includegraphics[scale=0.2]{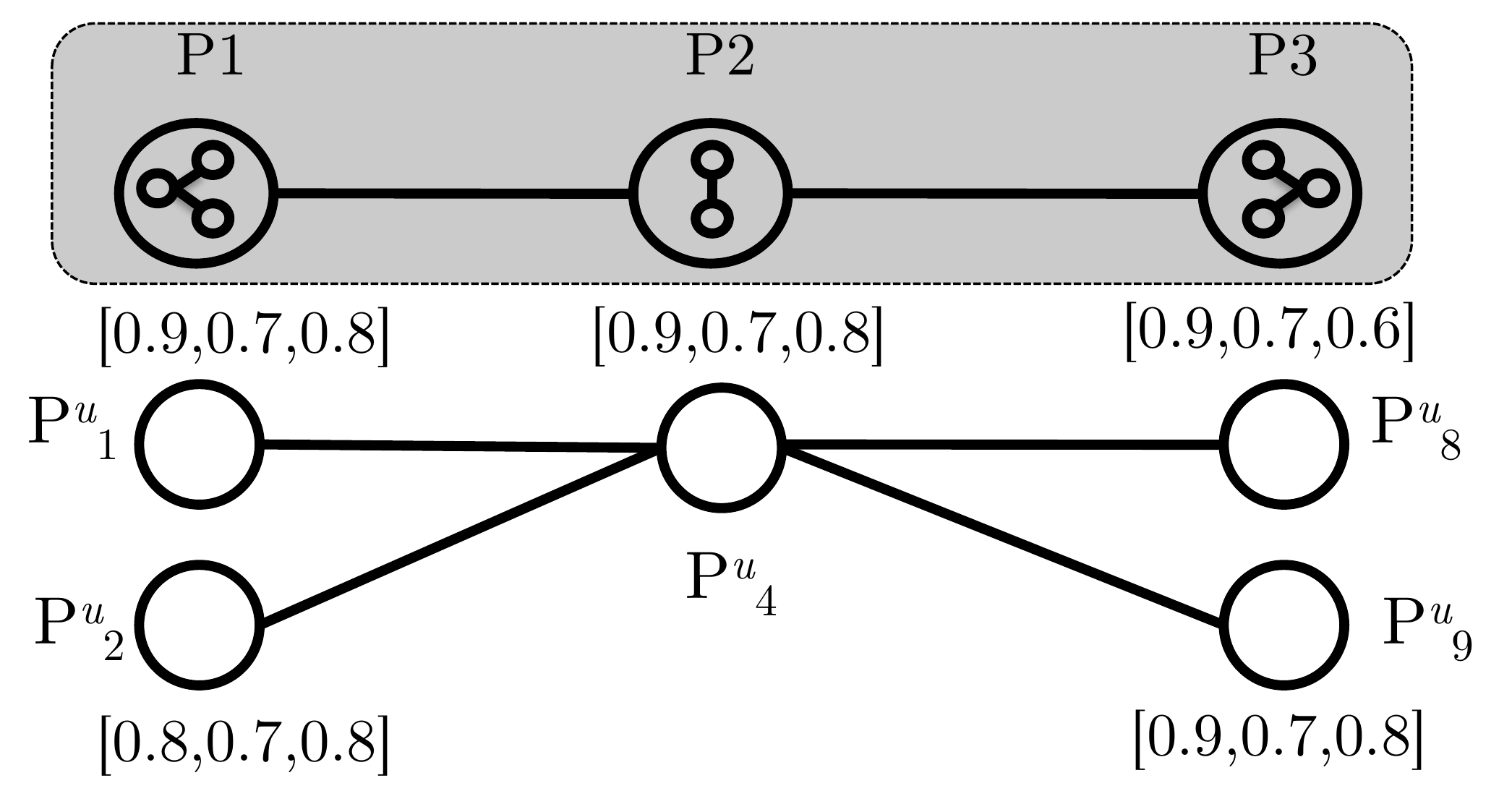}
\\
(e)
&
(f)
\\
\end{tabular}
\end{center}
\caption{(a) An example query and its decomposition, (b) k-partite graph construction, (c) reduction by structure, (d), (e), (f), reduction by upperbounds}
\label{fig:reduction}
\end{figure*}

\vspace{5pt}\noindent\textbf{Example: } \figref{fig:reduction}(a) shows an example of a query that is decomposed to three paths, where $P_2$ joins with $P_1$ and $P_3$. In  \figref{fig:reduction}(b), we show an example of the k-partite graph construction, by introducing links between pairs of path matches that satisfy the join conditions. Once the k-partite graph is constructed, it can be reduced by removing vertices that do not have any links to a partition that it should join with. Therefore, $P^u_3,P^u_5,P^u_6,P^u_7,P^u_{10}$ can be removed with all their links, resulting in the k-partite graph in  \figref{fig:reduction}(c). We can further apply reduction by upperbounds as shown in Figures \ref{fig:reduction}(d), (e), (f). In  \figref{fig:reduction}(d), each vertex is initialized by a partition perception vector that is all 1's except for the position of its own partition, which is initialized by the vertex's own weight. In this example, we consider weights of type $w_1$ only for simplicity. In the second step, each vertex updates its upperbounds based on values from its neighbors, leading to the perception vectors in  \figref{fig:reduction}(e). \figref{fig:reduction}(f) depicts the result of applying another iteration of the reduction algorithm, by performing one more pass of message exchange. Assuming that the input query probability threshold $\alpha=0.4$, we can see that $P^u_8$ can be removed from the graph along with its links. At this point, no further changes to the k-partite graph can take place, and we can proceed to the final result generation step.
}

\subsubsection{Finding Full Query Matches}
The final step of the online query processing algorithm is finding the full query matches. 
The algorithm starts from the matches of one path and progressively adds matches of joining paths, based on an initially determined join order.

\vspace{5pt}\noindent 
\textbf{Join order determination.}
In principle, the optimal join order could be determined by minimizing the size of the intermediate results, that is, the sum of the numbers of candidates after each step. To avoid the extra burden of this step, we add paths to the join order one at a time, based on the following heuristic:
\begin{myenumerate}
\item Choose the path with the largest number of nodes overlapping with the paths that already exist in the order.
\item In case of ties, choose the path with the largest number of join predicates with the existing paths.
\item In case of ties, choose the path with smallest cardinality (estimated as in path decomposition).
\end{myenumerate}
In general, a node on the new path can participate in multiple join predicates with existing nodes. However, when choosing  the first path in the order, the first two criteria are equal for all the paths, and we just use the third one.

\vspace{5pt}\noindent
\textbf{Finding matches.} Given the join order $\{P_1,\ldots,P_{|\mathcal{P}|}\}$, we use the reduced candidate k-partite graph to construct matches incrementally. The initial set of matches are the vertices in the partition corresponding to $P_1$. Each match $M_i$ up to path $P_i$ is extended to matches up to $P_{i+1}$ as follows. We first identify all paths $P_j$ with $j\leq i$ that join with $P_{i+1}$. For each vertex in $P_{i+1}$'s partition that has a link to the corresponding vertex in $M_i$ for each such $P_j$, we extend $M_i$ to a match up to $P_{i+1}$ by adding that vertex's candidate match. We discard $M_i$ if there is no such vertex, and only produce those extended matches  that have probability at least $\alpha$ and do not contain two nodes sharing a reference.

\subsection{Handling Correlations}
\label{sec:algo-correlations}
To handle edge existence correlations discussed in \secref{sec:modeling}, we replace the independent
edge existence probabilities $Pr((s_1,s_2).\boldsymbol{e})$ in all the equations with their corresponding conditional
probabilities
$Pr((s_1,s_2).\boldsymbol{e}|s_1.\boldsymbol{l_1},s_2.\boldsymbol{l_2})$.
Since the end point node labels are required to match the labels of
the corresponding nodes in the query, this conditional probability can be computed 
directly from the CPT in most of the equations. 
The only exceptions are the equations for $ppu(v,\sigma)$ and $fpu(v,\sigma)$
(\secref{sec:context}), where the existence probability of an edge is needed but one of the
end point node labels is not known.
Since those two functions are upper bounds, we simply modify the equations to find the maximum value
over all possible labels of $v$. Although this reduces the pruning ability of the context
information, we found it to have a negligible impact in our experimental evaluation.

However, more complex dependencies between node labels and edges' existence pose a bigger challenge
to efficient indexing. Given a path $P^u$, we can still compute its marginal probability
$Pr(P^u)$ for the purpose of indexing. However, $Pr_{le}$ is not decomposable in that case, and 
there is no easy way to correctly compute the joint probability of two (or more) paths during the
latter phases of the online algorithm without accessing the underlying PEG and thus defeating the purpose
of indexing. We leave a detailed exploration of indexing in presence of complex dependencies to
future work.

\section{Experimental Evaluation}
In this section, we present the results of a comprehensive experimental evaluation using our prototype implementation. Our implementation is written in Java and uses the disk-based graph database engine Neo4j for storing the probabilistic graph, and the key/value store KyotoCabinet to store the index as a B+ tree. We begin by presenting the index construction algorithm's performance in terms of both time and space, and then demonstrate online query performance by comparing it to various baselines. We further study the effect of the different pruning methods we proposed on reducing the search space, and the relationship between the search space size and different parameters. Finally, we report results on two real-world datasets from DBLP and IMDB. 
For the first set of experiments, we use synthetic graphs whose structure is generated according to the preferential attachment model \cite{barabasi:science99}. To generate node label probabilities, we first generate a set of random probabilities $p_1,\ldots,p_{|\Sigma|}$, which we then weigh by a zipf distribution, i.e., $p'_i= \frac{p_i}{i}$, to introduce skew. We normalize those to obtain final probabilities $p''_i= \frac{p'_i}{\sum_j{p'_j}}$, which are assigned to node labels randomly. Edge probabilities are generated analogously. To generate reference sets corresponding to entities, we randomly choose $k$ subsets of nodes from the graph, each of size $s$ nodes, and randomly assign $r$ pairs of nodes per group to the same reference set. That is, reference sets are of size 2, and the maximum size of a connected component is $s$. Probabilities of reference sets are generated randomly. We use merge functions that average the underlying distributions for both node attributes and edge existence. 
In our experiments, we use four settings with 50k, 100k, 500k, and 1m references, and a number of relations equal to $5\times$ the number of references in every setting. We set $k=\text{No. of references}/1000, s=r=4$. We associate probability distributions with $20\%$ of the references, relations, and reference sets unless otherwise stated. 
These settings result in probabilistic entity graphs of sizes (54k/292k), (108k/583k), (540k/ 2.95m) and (1.08m/5.88m) nodes/edges, respectively. Synthetic experiments are performed on an Amazon EC2 instance with a Linux operating system,  8 core processors, 117 GB of RAM and 2 TB of instance storage. The realworld experiment is performed on a Linux machine with two 2.66 GHz quad-core processors with hyper-threading, 48 GB of RAM, and a 1TB 7200 RPM disk drive.
\subsection{Offline Phase Performance}
We first compare performance of the offline phase for maximum
index path lengths $L = 1,2,3$.

\begin{figure*}
\begin{center}
\begin{tabular}{@{\extracolsep{-10pt}}c c}
\includegraphics[scale=0.45]{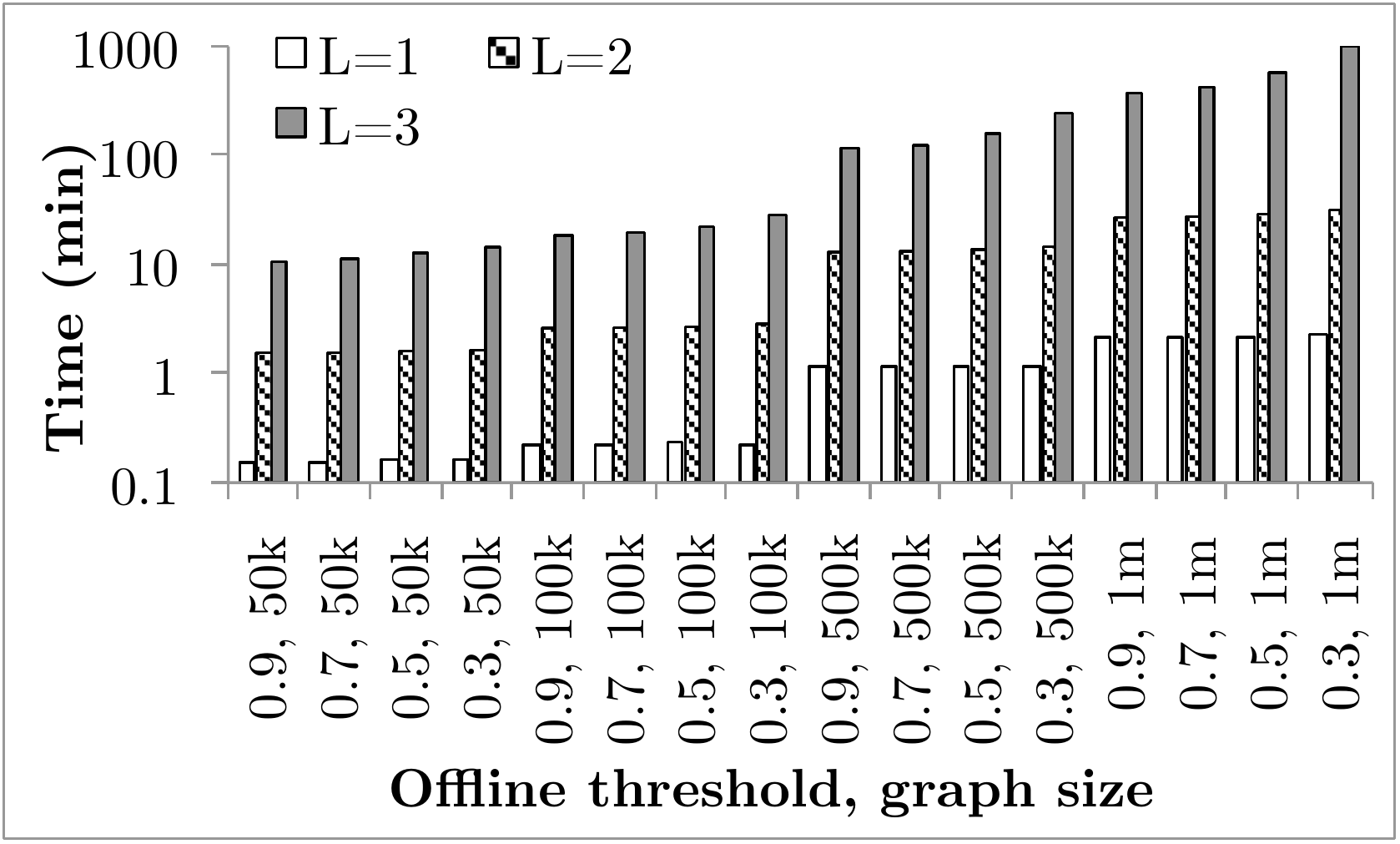}
&
\includegraphics[scale=0.45]{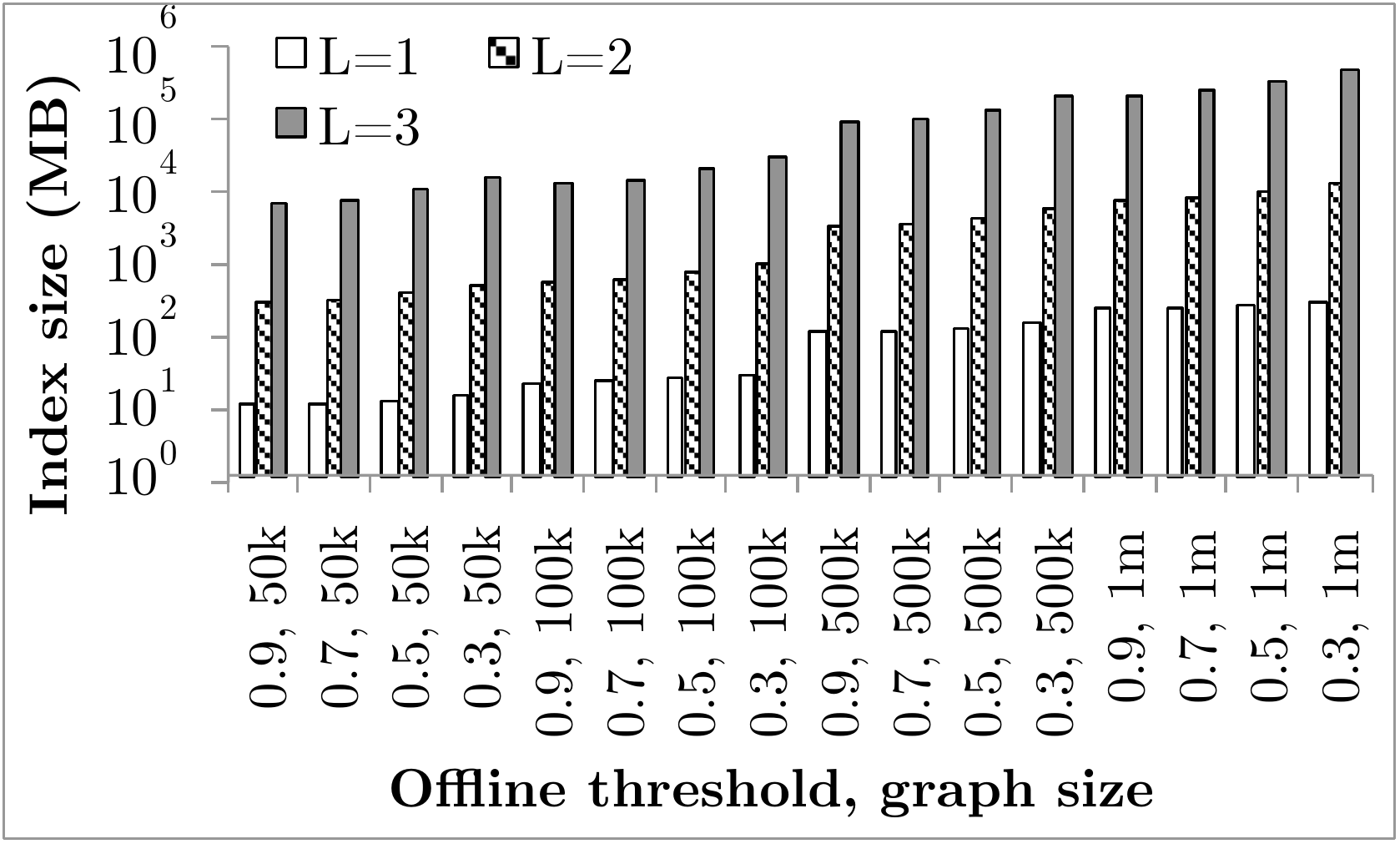}
\\
(a) & (b)\\
\includegraphics[scale=0.45]{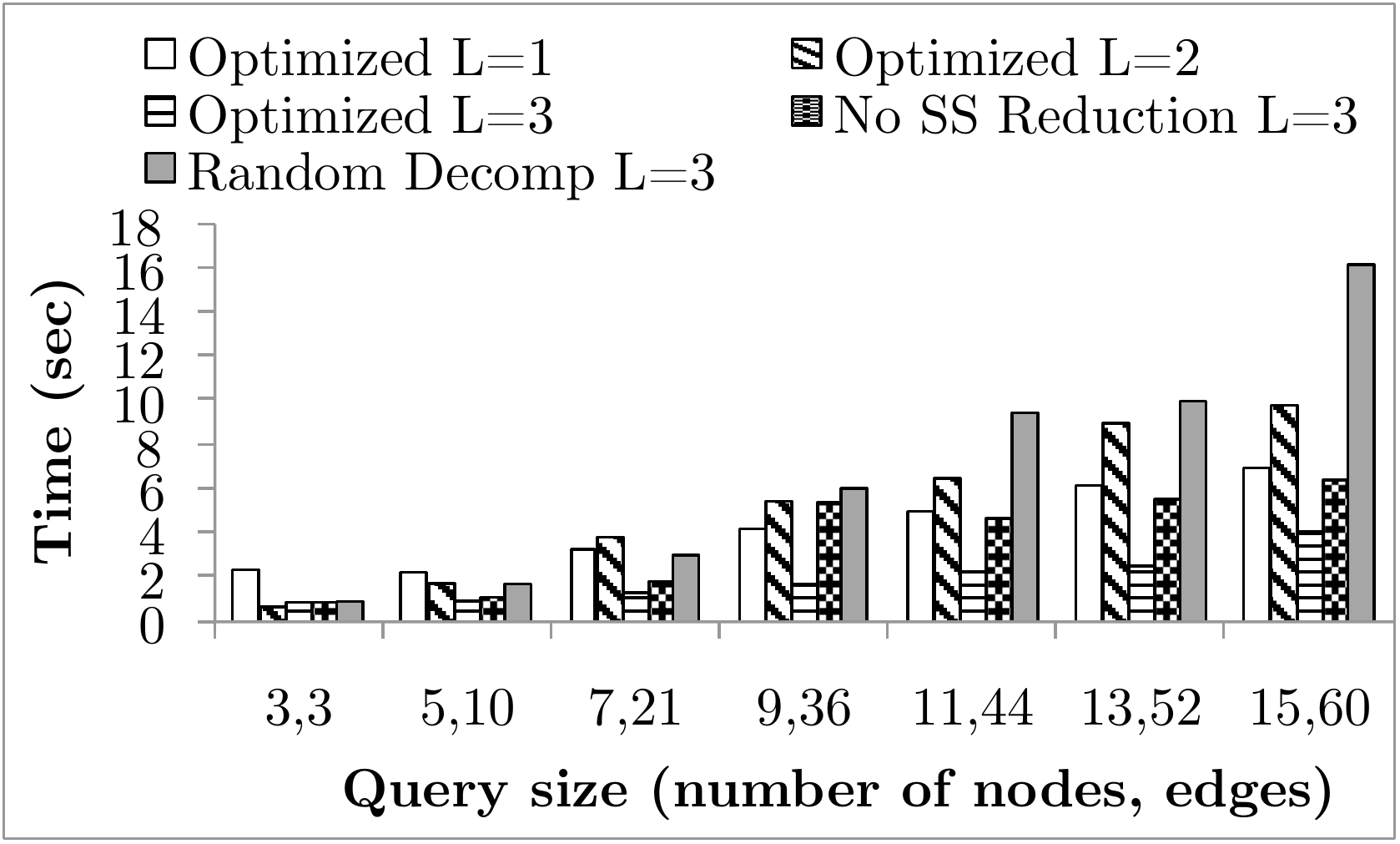}
&
\includegraphics[scale=0.45]{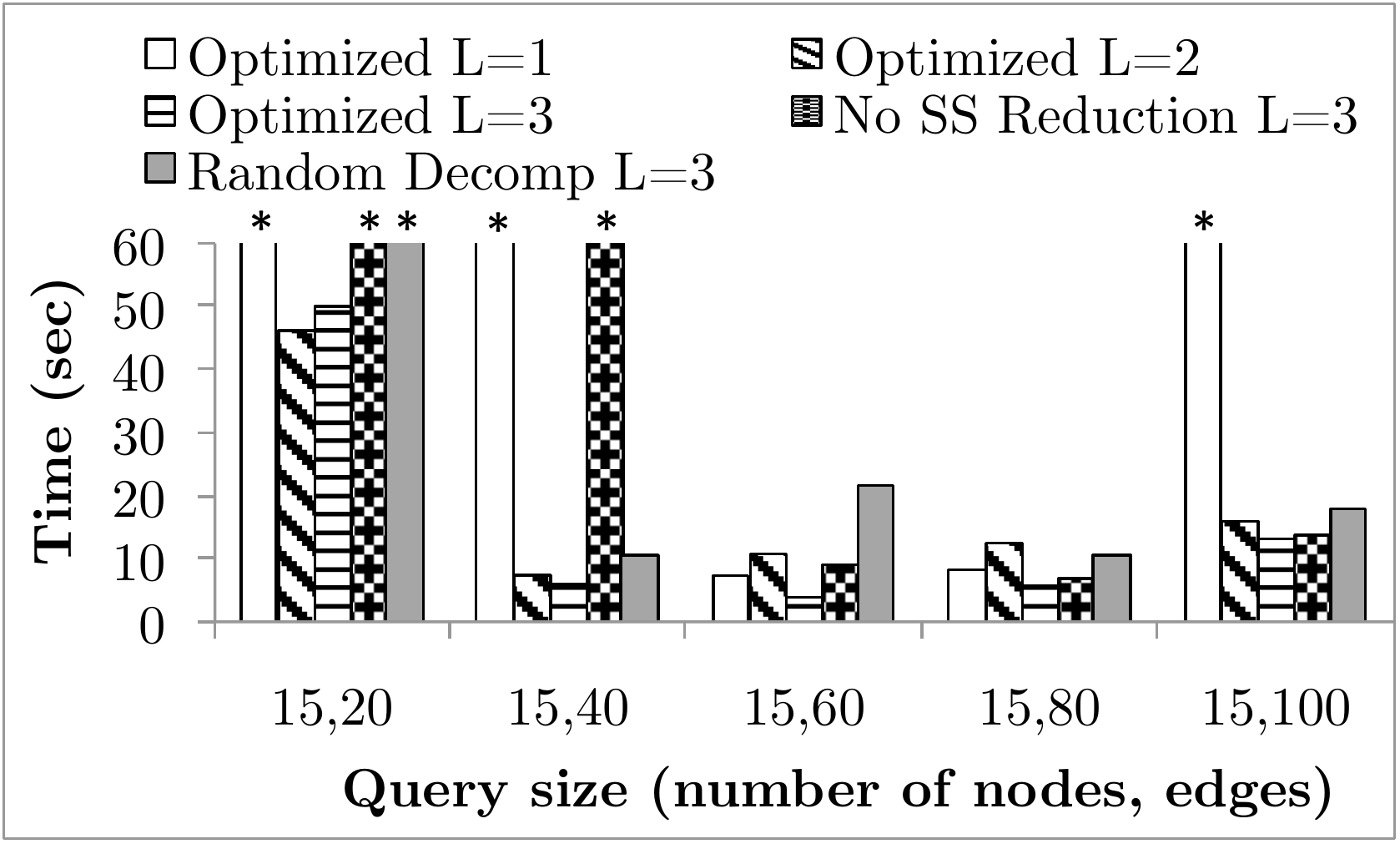}
\\
(c) & (d) \\
\includegraphics[scale=0.45]{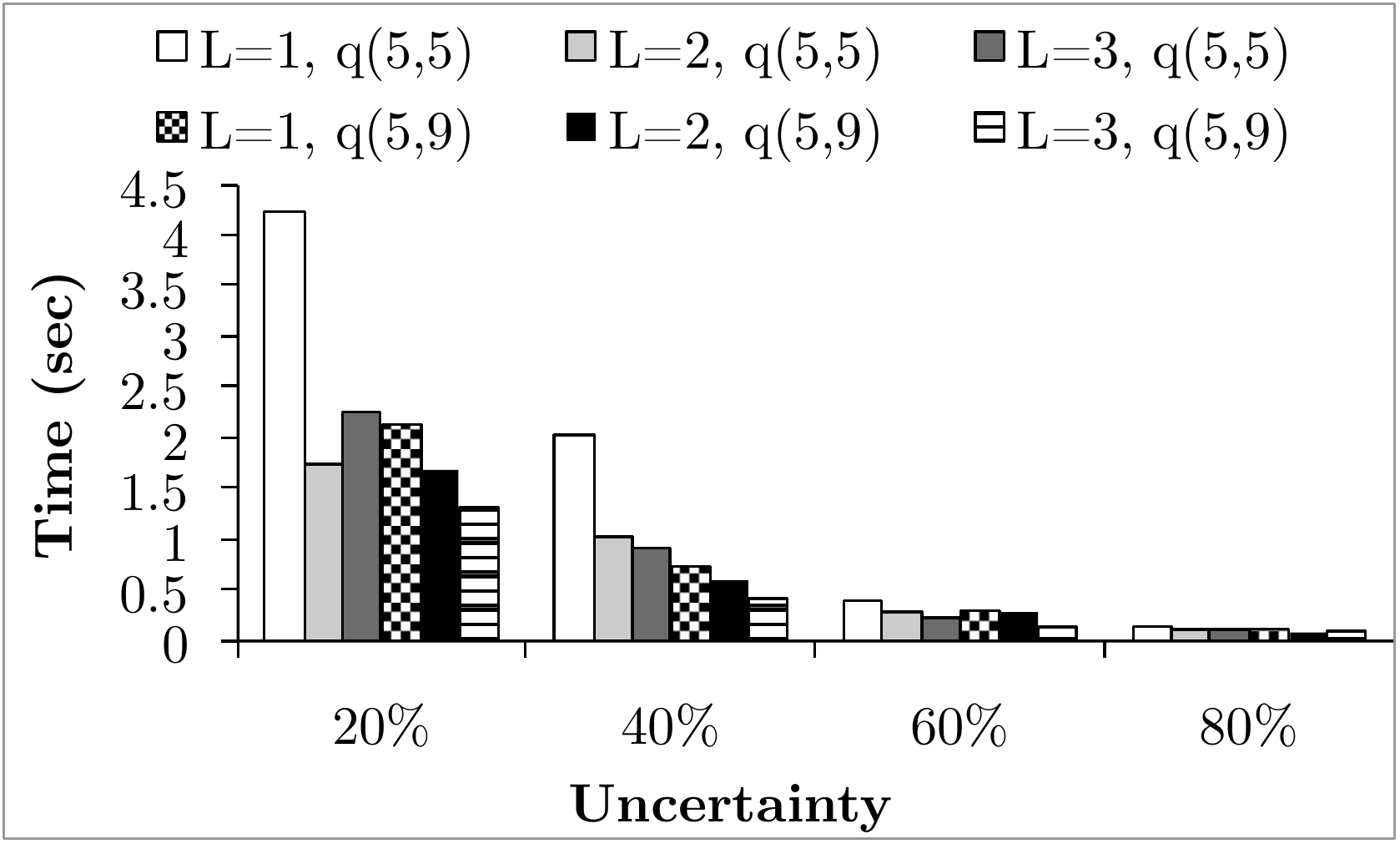}
&
\includegraphics[scale=0.45]{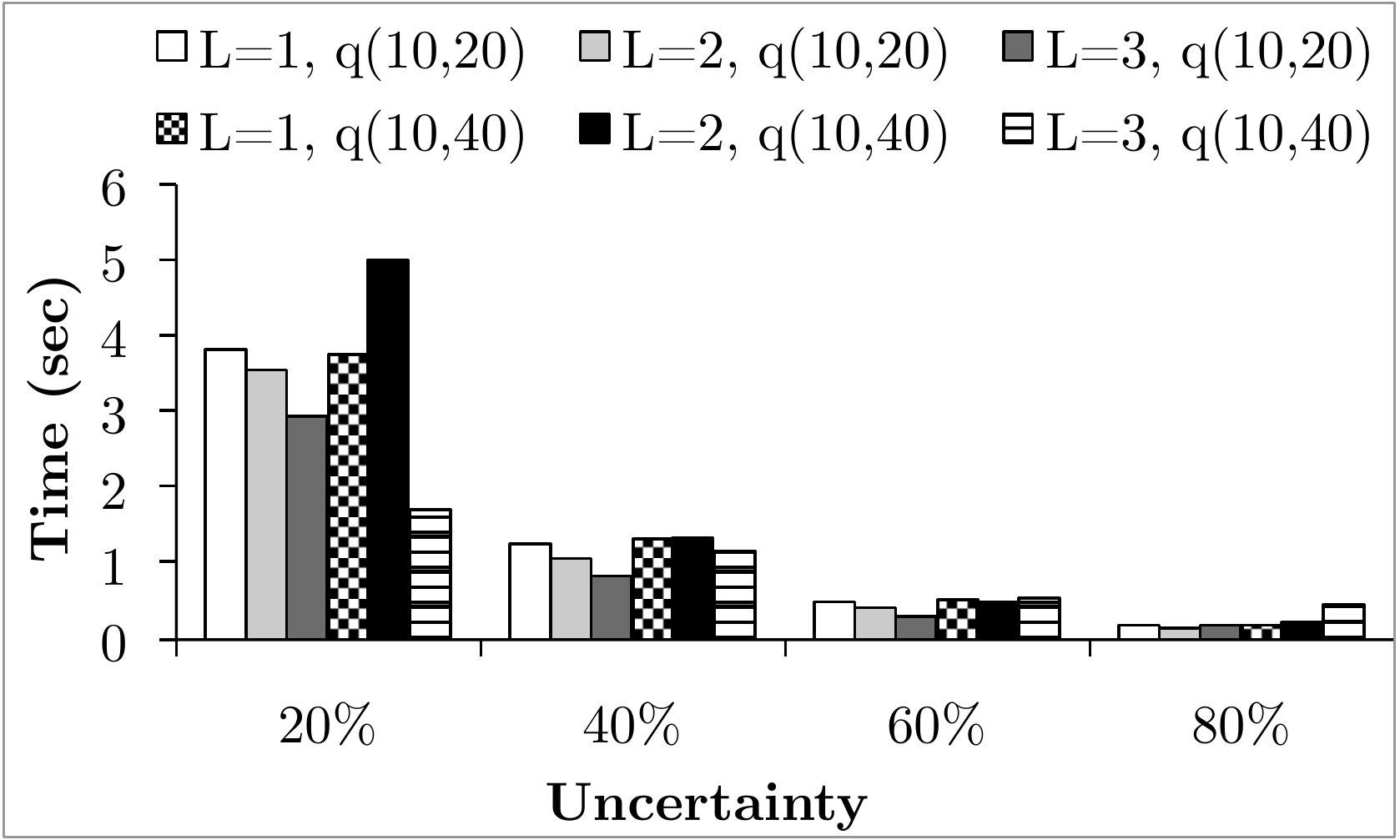}
\\
(e) & (f) \\
\end{tabular}
\end{center}
\caption{(a),(b) Offline phase performance, (c) varying query size, (d) varying query density, (e), (f), varying degree of uncertainty for queries with 5 and 10 nodes, respectively. A * above a bar indicates that the query did not finish in the allocated time (15 minutes), or the process ran out of memory.}
\label{fig:results1}
\end{figure*}

\paragrph{Running Time:}
We first study the running time performance of the entire offline
phase, which  includes calculating the entity graph component
probabilities, building the path index, and calculating context
information. \figref{fig:results1}(a) shows the running time when
varying  both the graph size and the index lowerbound probability threshold~$\beta$. 
The offline phase running time at $L=2$ is between 10 and 14 times that at $L=1$, and at $L=3$ it is between 7 to 30 times that of $L=2$. Also, as the graph size increases, running time increases by a factor  less than the graph size increase factor. For example, although the 1m graph is $20$ times larger than the 50k graph, the running time 
increases by a factor of 14 on average at $L=1$, 18 on average at $L=2$, and 46 on average at $L=3$. This is due to higher memory buffer utilization for larger graphs. 

\paragrph{Path Index Size:}
We next compare the path index size, varying the graph size and index threshold as before. Results in \figref{fig:results1}(b) show that index sizes at $L=2$ are 32 times larger than those at $L=1$ on average, and index sizes at $L=3$ are 28 times larger than those at $L=2$ on average. Index size increases at the same rate as the graph size at $L=1$, and faster than the increase in the graph size at $L=2$, e.g., the index size at 1m is 20 times larger than that of 50k on average at $L=1$ and 25 times on average at $L=2$. This is because indexes at $L=1$ increase linearly with graph size, while at $L=2$ the index size increases quadratically. The same trend applies at $L=3$ as its size increases cubically.

\subsection{Online Phase Performance}
We now study different performance aspects of the online phase.
\subsubsection{Online running time}
We first compare the running time of our proposed algorithm to a range of baselines, using different input query sizes. We use the following algorithms and parameters:
\begin{myenumerate}
\item \textbf{Optimized:} This refers to our proposed approach with all the proposed optimizations. We use path lengths $L=1,2,3$.
\item \textbf{Random decomposition:} This is a variant of our proposed approach that does not employ the proposed query decomposition algorithm. In this baseline, we use random query decomposition instead of SET COVER, and when determining the path join order, we sort the paths according to their number of path index matches only, without taking into account the number of node intersections, number of predicates, path degree or path density. We set $L=3$ for this baseline.
\item \textbf{No search space reduction:} This approach uses our optimized method, but without the joint search space reduction using the k-partite graph representation, and goes directly to generating final results after constructing the candidate and relative candidate lists. We set $L=3$ for this baseline.
\item \textbf{SQL:} We implement our queries using SQL and run them on top of MySQL database. We run SQL on the 100k nodes dataset using a query with 5 nodes and 7 edges and a query threshold of 0.7. While our approach can answer this query in less than a second, SQL never finishes it in a month. Therefore, we do not report any other SQL-based performance metrics.
\end{myenumerate}

\paragrph{Varying input query size:}
In this experiment, we study the running time performance of Optimized ($L=1,2,3$), Random Decomp and No SS Reduction for varying query size. We use the 100k dataset and a query threshold of $0.7$. \figref{fig:results1}(c) shows running times for 7 different query sizes between q(3,3) and q(15,60), where q($n$,$m$) denotes a query with $n$ nodes and $m$ edges, averaged over five randomly generated queries per size. A query of $n$ nodes has $4 \times n$ edges, unless the maximum number of edges for the query is less than $4 \times n$, in which case, we use the maximum possible number of edges. Our approach at $L=3$ always outperforms $L=1,2$ and both of Random Decomp and No SS Reduction.  For smaller queries (with 3 and 5 nodes), $L=2$ outperforms $L=1$, but it does not for the larger ones. The reason is that $L=1$ has an advantage with querying the path index, as it returns a lower number of matches than both $L=2,3$, and at the same time, $L=3$ has an advantage with context-based pruning, as higher path lengths have richer context information. At $L=2$ the pruning performed with context information does not alleviate the processing needed for the larger number of matches returned from the path index, especially with larger query sizes. However, as we show in further experiments, $L=2$ outperforms $L=1$ when the input graph has higher degree of uncertainty, even for larger query sizes, and also sometimes outperforms $L=1$ in extreme cases, such as queries with a very large number of results, or with a very large number of nodes and edges, or very large input graphs (e.g., (500k, 2.5m) and (1m, 5m) nodes, edges). Therefore, even though $L=2$ sometimes does not perform as well as $L=1,3$, it may be used as a compromise that does not take as much time and space as $L=3$ in building its index, and still has an acceptable performance in extreme cases where $L=1$ may not succeed.

\paragrph{Varying input query density:}
In this experiment, we study the running time performance of Optimized
($L=1,2,3$), Random Decomp and No SS Reduction for varying the input
query density. We use the 100k dataset and a query threshold of
$0.7$. \figref{fig:results1}(d) shows running times for 5 different
densities, by using queries with 15 nodes and between 20 and 100 edges. Each result is the average over five randomly generated queries with the corresponding size. Again, our approach at $L=3$ always outperforms $L=1,2$ and both of Random Decomp and No SS Reduction.  $L=1$ runs out of memory at the query q(15,20) due to the large number of matches of that query (because it is very sparse). Therefore, we do not show its running time. Furthermore, there are configurations which have at least one run of the five runs whose execution time exceeded the maximum time allowed of 15 minutes. Those configurations are $L=1$ at q(15,40), q(15,100), No SS Reduction at q(15,20), q(15,100), and Random Decomp at q(15,20).

\paragrph{Varying input graph degree of uncertainty:}
In this experiment, we study the effect of the degree of uncertainty in the PEG on the running time of our proposed approach, by varying the number of uncertain nodes and edges from $20\%$ to $100\%$. We use query sizes q(5,5) and q(5,9) (in \figref{fig:results1}(e)), and q(10,20) and q(10,40) (in \figref{fig:results1}(f)), with a query threshold of $0.7$. As we can see, $L=3$ always outperforms $L=1,2$, while $L=2$ outperforms $L=1$ for all degrees of uncertainty larger than $20\%$.
\begin{figure*}
\begin{center}
\begin{tabular}{@{\extracolsep{-10pt}}c c}
\includegraphics[scale=0.45]{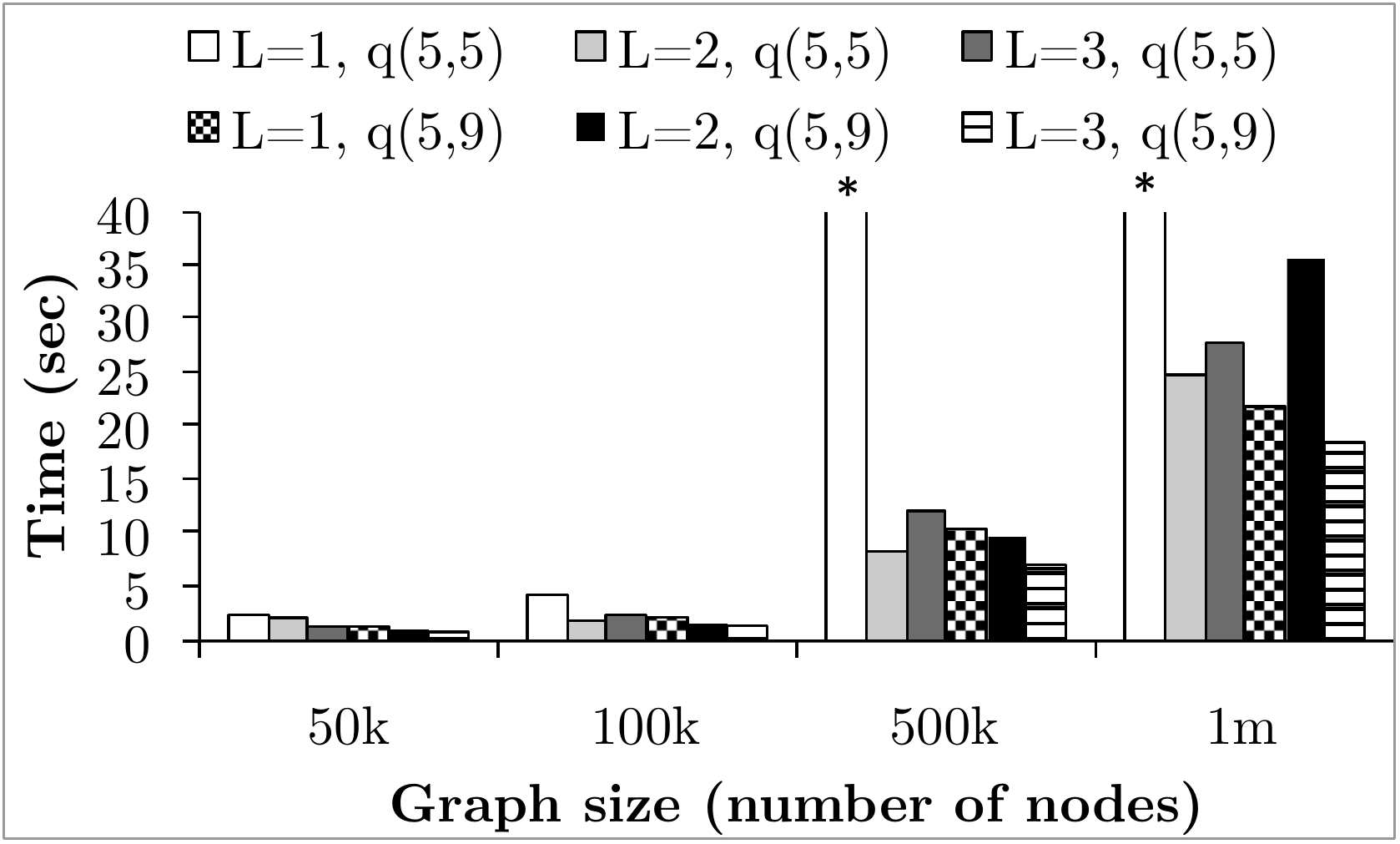}
&
\includegraphics[scale=0.45]{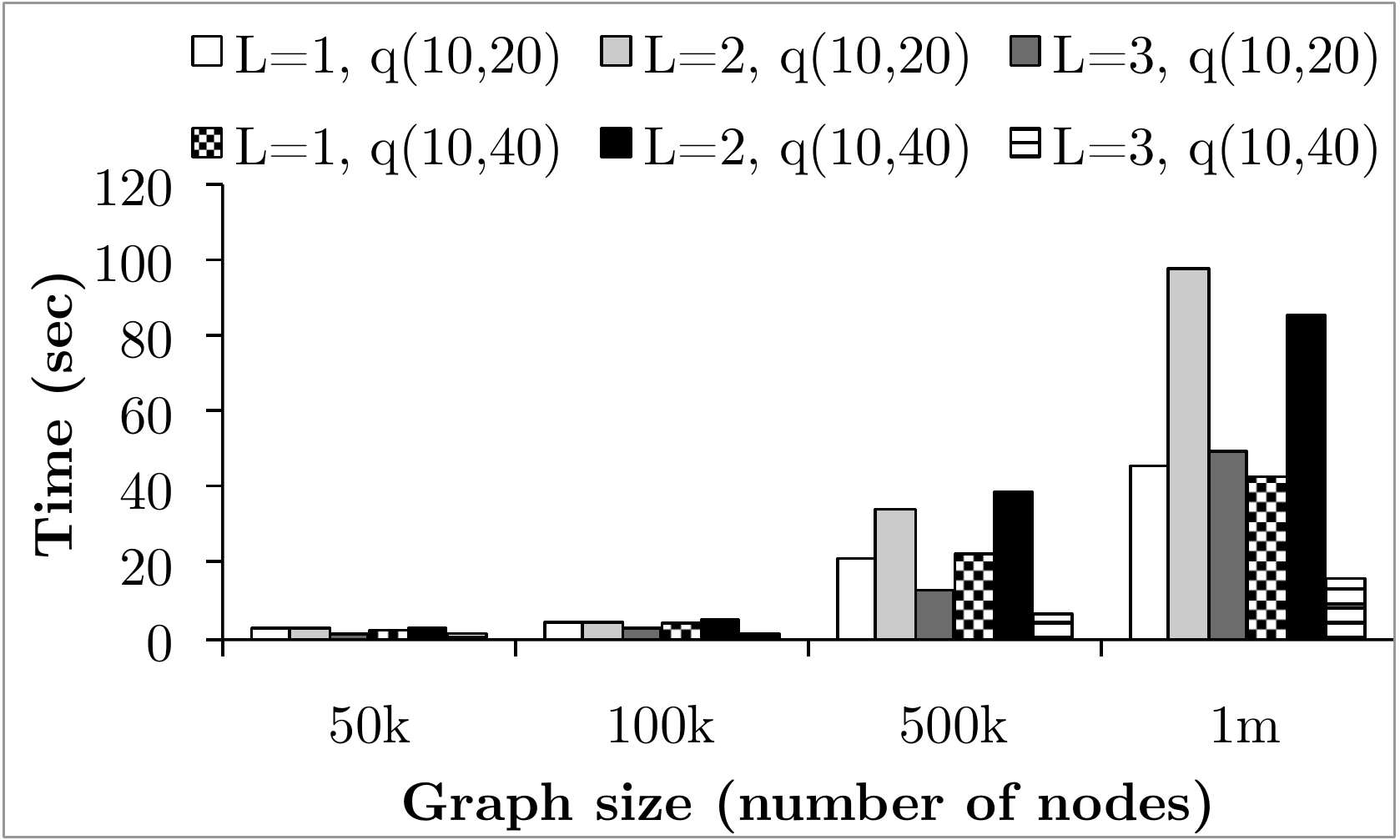}
\\
(a) & (b) \\
\includegraphics[scale=0.45]{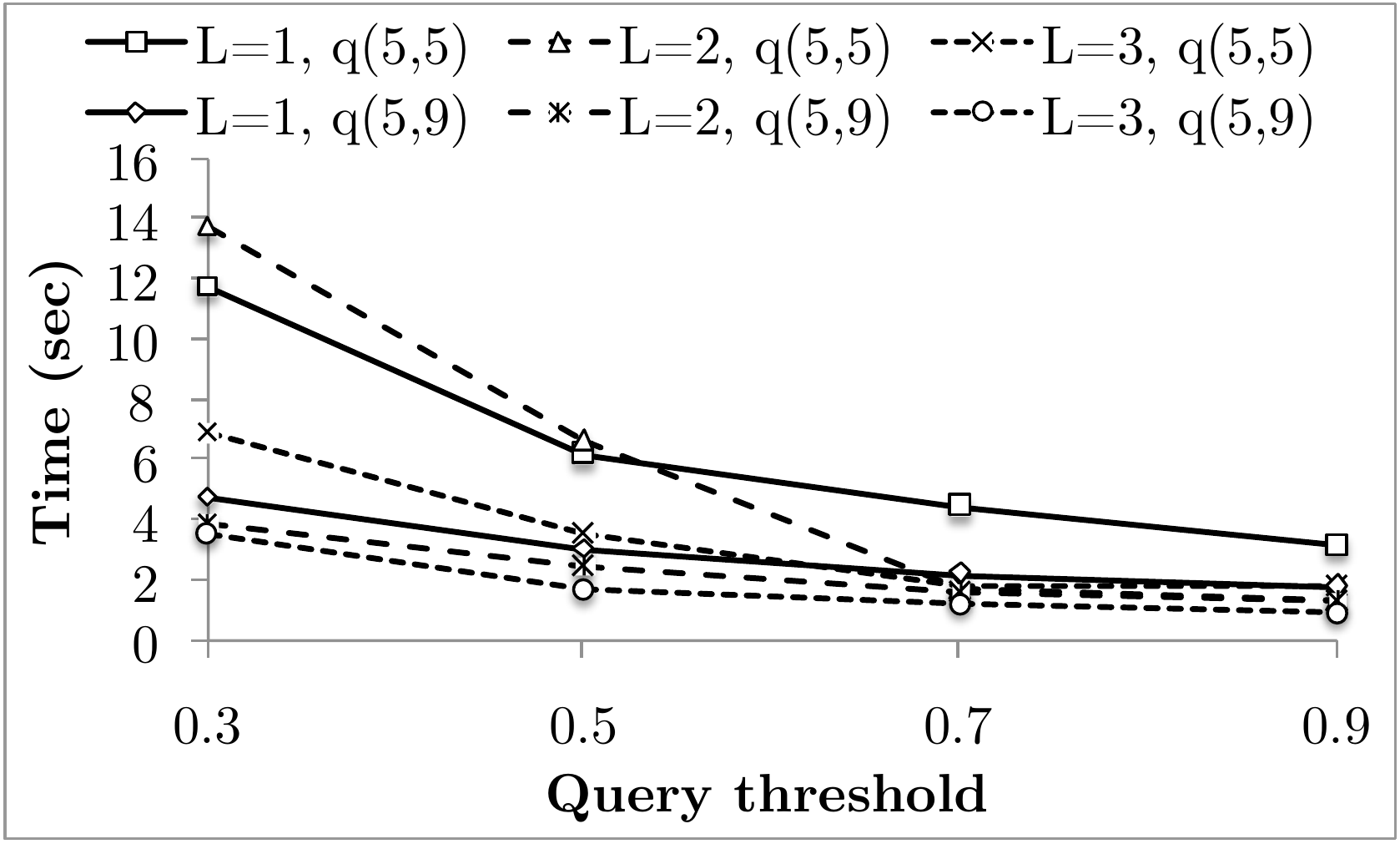}
&
\includegraphics[scale=0.45]{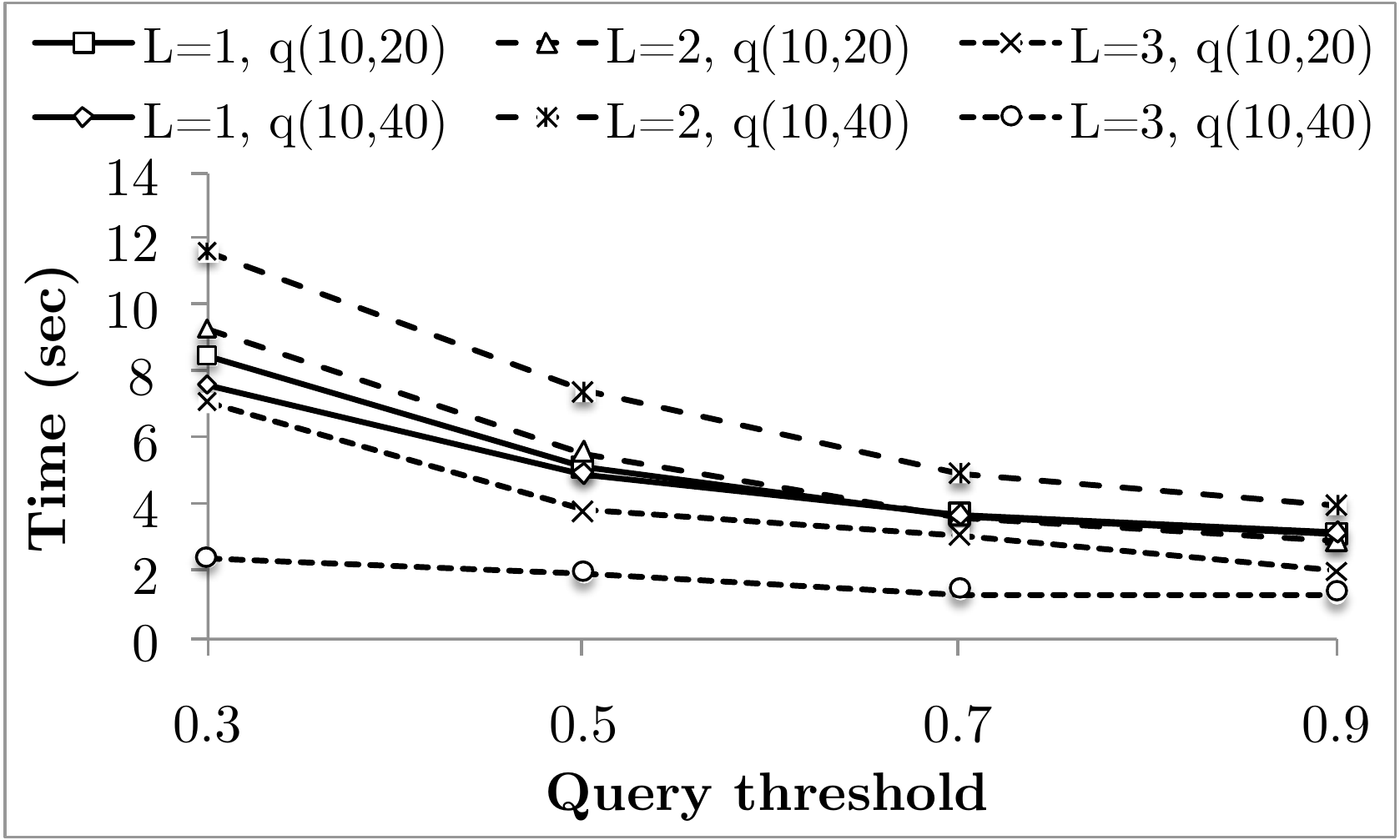}
\\
(c) & (d) \\
\includegraphics[scale=0.45]{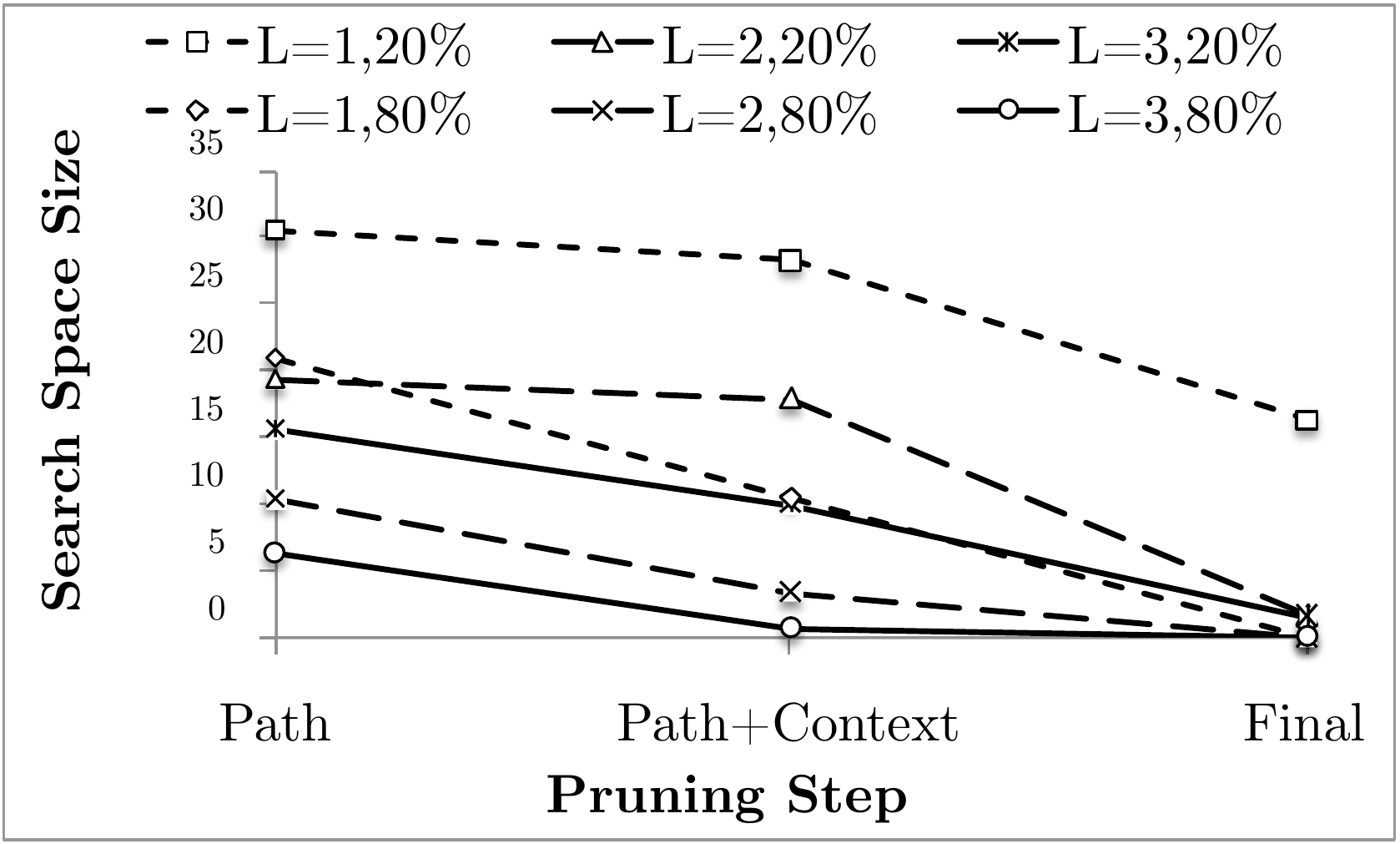}
&
\includegraphics[scale=0.45]{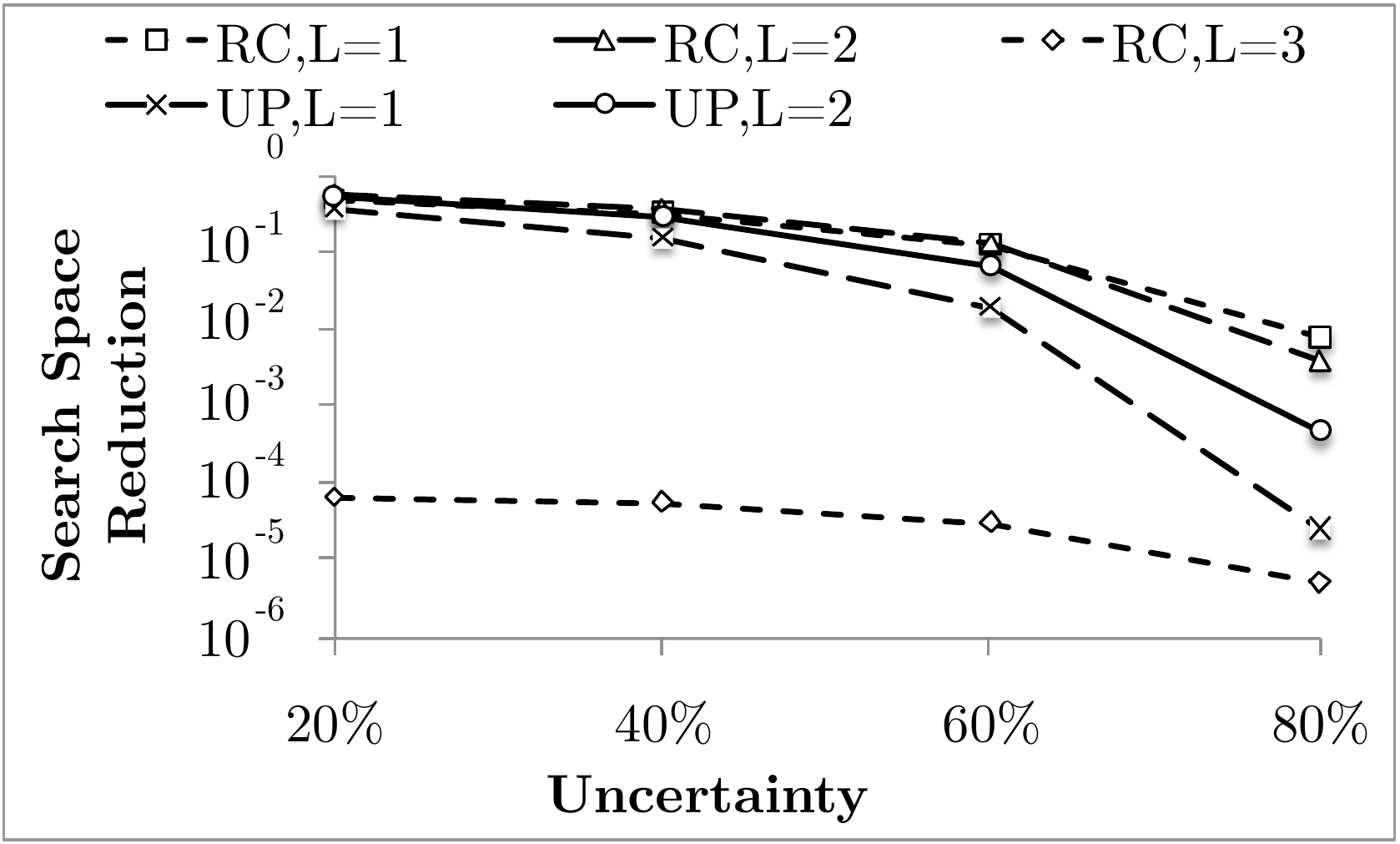}
\\
(e) & (f)\\
\includegraphics[scale=0.45]{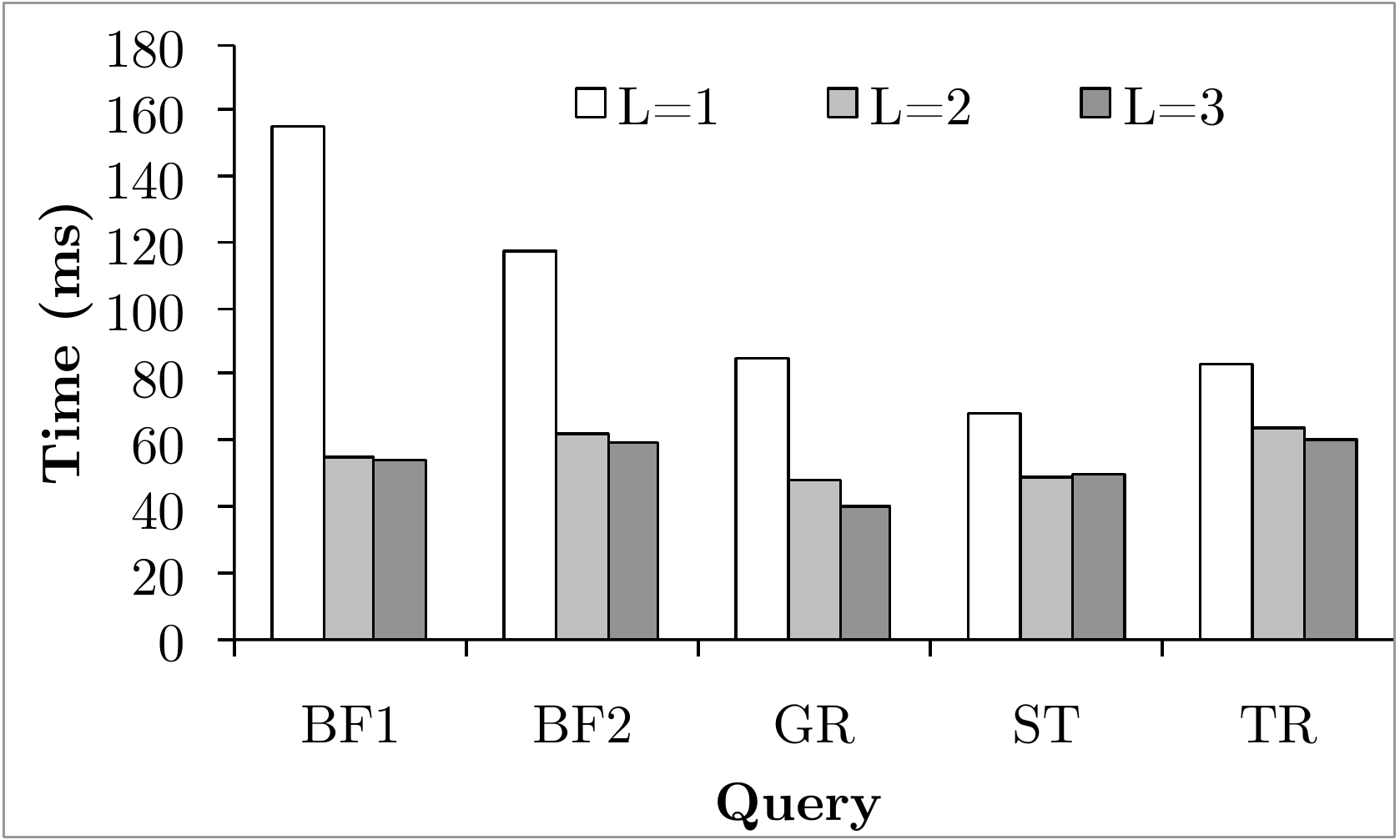}
&
\includegraphics[scale=0.45]{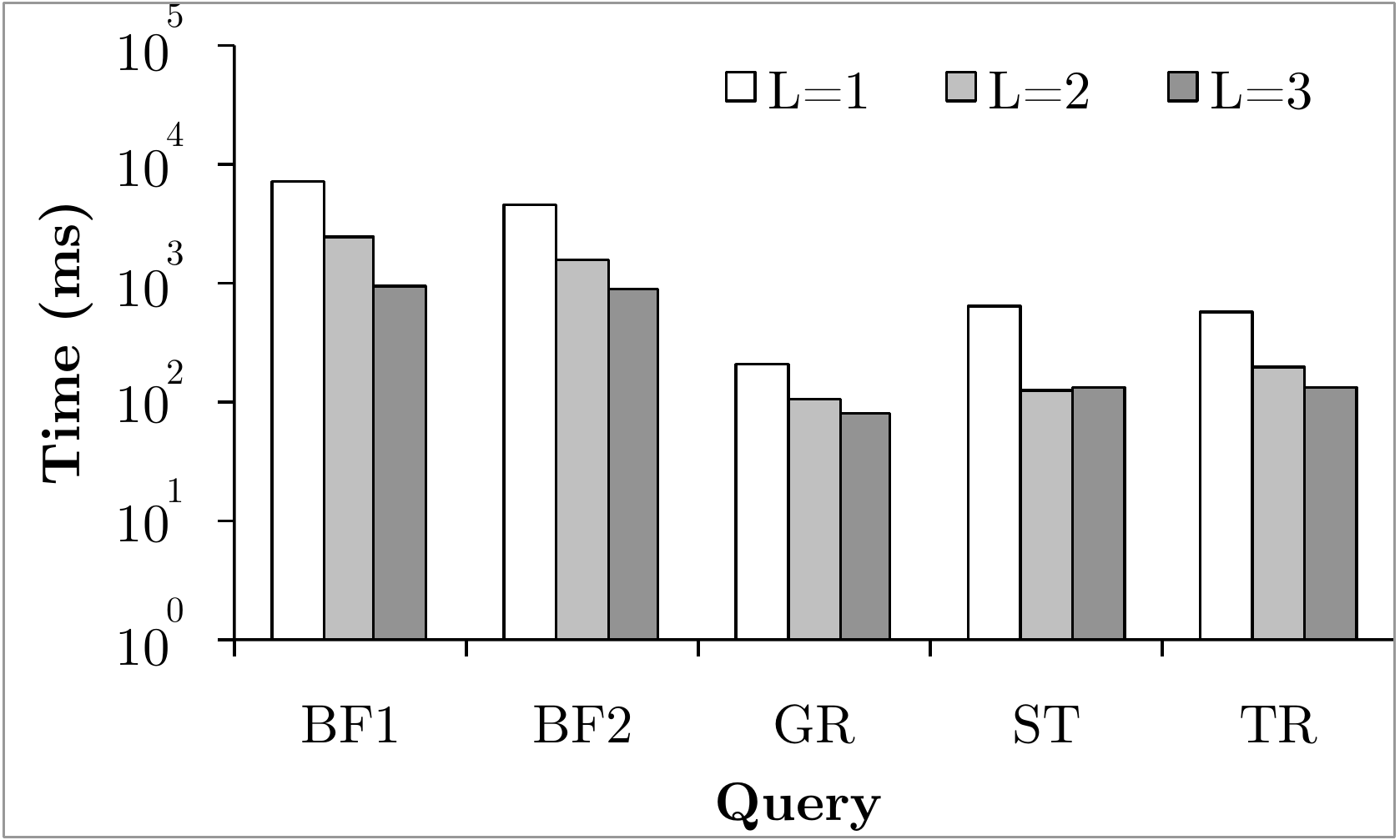}

\\
(g) & (h) \\
\end{tabular}
\end{center}
\vspace{-15pt}
\caption{(a),(b) Varying input graph size for queries with 5 and 10 nodes, respectively, (c), (d) varying input query threshold for queries with 5 and 10 nodes, respectively, (e),(f) search space experiments, (g), (h) performance on the DBLP, and IMDB real-world datasets, respectively. A * above a bar indicates that the query did not finish in the allocated time (15 minutes), or the process ran out of memory.}
\label{fig:results2}
\end{figure*}

\paragrph{Varying input graph size:}
In this experiment, we study the performance of our proposed approach
for all four input graph size settings, corresponding to graphs whose
number of edges varies between 300 thousand and 6 million. We use
query sizes q(5,5) and q(5,9) (in \figref{fig:results2}(a)), and
q(10,20) and q(10,40) (in \figref{fig:results2}(b)), with a query
threshold of $0.7$. With query q(5,5), $L=1$ runs out of memory at both 500k and 1m due to the high number of matches, while $L=2,3$ finishes normally in those cases. Otherwise, $L=3$  outperforms $L=1,2$ in most cases.

\paragrph{Varying input query threshold:} We vary the query threshold between 0.3 and 0.9. We use queries of size q(5,5), q(5,9) (in \figref{fig:results2}(c)) and q(10,20), q(10,40) (in \figref{fig:results2}(d)), using the 100k dataset. The performance improves for all path lengths with increasing threshold, but at the same time, the performance of lower path lengths is the most sensitive to the change in the threshold, indicating that higher path lengths are the most stable with respect to such a parameter.

\subsubsection{Search Space Performance}
In this set of experiments, we study the search space performance, measured as the product of the candidate list sizes, and its reduction throughout different steps of our proposed method, under different circumstances.

\paragrph{Search Space Progression:}
In this experiment, we study the progression of the search space size throughout the main steps of our online querying algorithm. The results are depicted in \figref{fig:results2}(e). 
The first step (labeled Path) refers to the search space size resulting from querying the path index. 
The second step (labeled Path+Context) is the size of the search space after pruning based on context information, that is, node-based neighborhood information, path neighbors and path cycle, as discussed in Section~\ref{finding-path-candidates}. 
The last step (labeled Final) refers to the final search space size after applying the mutual search space reduction using the k-partite graph representation (Section~\ref{sec:join-reduction}). We use a randomly generated query of size q(5,7) with query threshold of 0.7 over two 100k datasets, one with $20\%$ uncertainty, and the other with $80\%$ uncertainty. \figref{fig:results2}(e) shows the performance of our approach (in log scale) using the three path lengths of $L=1,2,3$. As we can see, the mutual search space reduction step (Final) achieves effective reduction for all path lengths, although it is more effective with shorter path lengths. 
This is due to the fact that decompositions with shorter paths take into account information from smaller neighborhoods, and thus benefit more from distant information obtained via message passing. 
In contrast, the previous step (Path+Context) is most effective for longer paths, as those provide more context  information for pruning. Also, generally, higher degree of uncertainty results in smaller search spaces, because more paths are pruned at every step compared to lower degrees of uncertainty. Finally, we can see that overall, the final search space for longer paths is much smaller than that for shorter ones, which emphasizes the effectiveness of higher values of $L$ in producing much smaller search spaces: \emph{14 orders of magnitude smaller, comparing $L=3$ to $L=1$ at $20\%$}.

\paragrph{Joint Search Space Reduction Performance:}
In this experiment, we study the mutual search space reduction step (Section~\ref{sec:join-reduction}) in more detail, taking a closer look at the performance of both reduction methods: reduction by structure (ST), and reduction by upperbounds (UP). We use  graphs of size 100k, a query that is a cycle with 5 nodes and 5 edges, and threshold 0.1. We have chosen a cycle query because it has a high diameter, thus illustrating the performance of information exchange using both reduction methods along the edges. For each method, we measure its reduction by dividing its resulting search space size by the initial search space size immediately before the reduction algorithm starts. Of course, since reduction by upperbounds is performed after reduction by structure, it will always perform higher reduction, but we are interested in its contribution to the overall reduction, and how it is affected by different parameters. \figref{fig:results2}(f) shows the search space reduction for both ST and UP using three different path lengths 1, 2, 3 over graphs whose degrees of uncertainty vary from $20\%$ to $80\%$. We do not show the case (UP,L=3) because the algorithm terminated (i.e., no further changes took place) before reduction by upperbounds already. As we can see, the effect of both reduction methods increases with the degree of uncertainty in the graph, again because more paths can be pruned. However, particularly, the effectiveness of UP increases with increased degree of uncertainty, as increased uncertainty often results in tighter upperbounds.
Finally, we observe that reduction by upperbounds is more effective with shorter path lengths, as those obtain more additional information during message passing, while longer ones have already exploited part of this information during context based pruning and reduction by structure.  

\subsection{Performance on Real-world Data}
\begin{figure}
\begin{center}
\includegraphics[scale= 0.15]{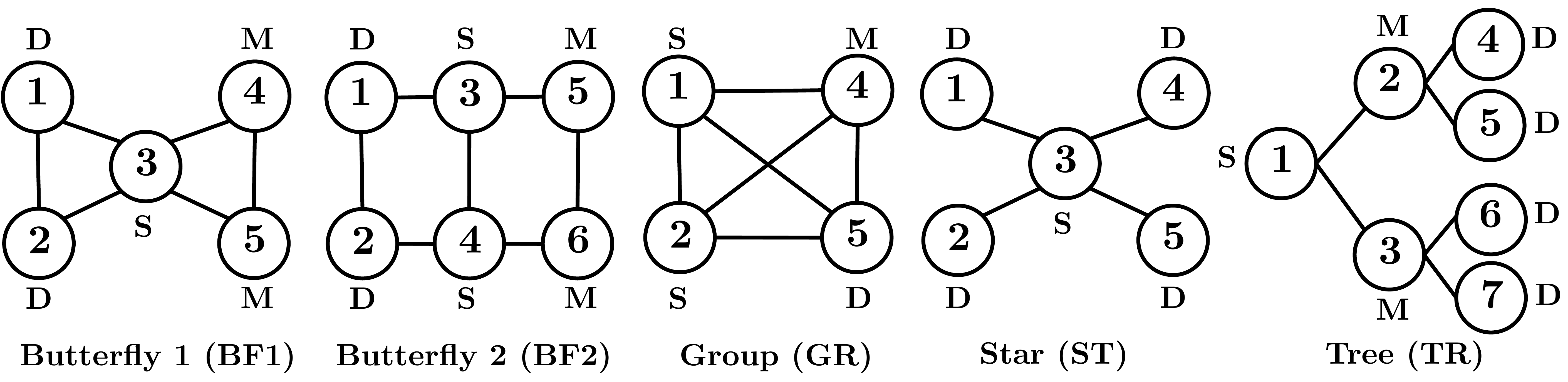}
\end{center}
\vspace{-15pt}
\caption{Pattern queries for real-world datasets.}
\vspace{-15pt}
\label{fig:realworld-queries}
\end{figure}
In this subsection, we show our experimental results on two real-world
datasets, DBLP and IMDB. We use correlated edge and label
probabilities with DBLP, and independent edge probabilities with
IMDB. 
For the DBLP network, we extract the ``author collaboration'' graph. The nodes of the graph represent authors, the edges represent collaboration relationships. We annotate the collaboration graph with probabilistic data to capture different types of uncertainties. For every author, we assign a probability distribution over the areas that she/he is interested in, which can be Databases, Machine Learning, or Software Engineering. We extract this information by counting the author's relative contribution in each area's conferences. For example, SIGMOD, VLDB, and ICDE count towards Database interests, while ICSE, FSE and ICSM count towards Software Engineering interests, and so on. 
To obtain the edge existence probability for a pair of authors, we first generate a base
probability between 0.5 and 1 depending on the number of
collaborations between them. If the authors' research interests as given by the
node labels are the same, the conditional edge existence probability is
the base probability~$p$, else, it is 
$0.8\cdot p$.
We create a reference set for every pair of authors whose names have normalized string similarity score above 0.9. The resulting graph has 16.8k nodes and 40.3k edges. We run probabilistic subgraph pattern matching using the collaboration patterns shown in \figref{fig:realworld-queries}  with a query threshold of 0.1. Running times of the online phase using $L=1,2,3$ are shown in \figref{fig:results2}(g). As we can see, $L=3$ outperforms $L=2$, which in turn outperforms $L=1$,  for all queries except the tree query.

The IMDB network is a ``co-starring'' graph, that is,  nodes are
actors, and edges are co-starring relationships between actors. We
use Drama, Comedy, Family and Action movies from the IMDB dataset, and create a co-starring edge between the two main stars of each movie. Standard statistical prediction methods are used to introduce probabilities to the network, where node attribute uncertainty are obtained from the distribution over movie genres an actor participates in, co-starring edge probabilities are obtained from the number of times two actors co-star together, and identity uncertainty is obtained from similarities in actor names, which may have occurred from duplicates or misspellings. The size of this network is 90,612 nodes and 936,308 edges. We use the same query structure of queries depicted in \figref{fig:realworld-queries}, with co-starring edges linking nodes of the same genre, i.e., each query has the same label for its nodes and the label is randomly generated. The input probability threshold $\alpha$ is $0.1$. Results are shown in \figref{fig:results2}(h), again we observe that $L=3$ outperforms $L=2$, which in turn outperforms $L=1$.

\section{Related Work}
\coloredcomment{\color{red} (in the light of space constraints, are there some references that we could drop, especially from the general lists in the beginning?)}

Although many research studies have addressed the problems of representing and querying uncertain and probabilistic data, 
e.g., \cite{dalvi:vldbj07,sen:vldbj09,kanagal:sigmod09}, the area of uncertain graph data processing is still new and gaining more interest recently. Research in uncertain graph databases has covered different areas such as finding shortest paths, reliable subgraphs, mining frequent patterns, and answering graph queries, e.g., \cite{potamias:pvldb10,jin:pvldb11,jin:kdd11,hintsanen:dmkd08,zou:icde10,papapetrou:edbt11,zou:tkde10,chen:tkde10,yuan:pvldb11,yuan:vldb12}. 


Udrea et al., \cite{udrea:IRI06} propose precise semantics for probabilistic RDF graphs formed by associating probabilities to triplets, calling them quadruples. They propose algorithms for answering queries consisting of one quadruple with one variable at most.  Huang el al., \cite{huang:wise09} propose algorithms for query processing over probabilistic RDF graphs with edge uncertainty only. Lian et al., \cite{lian:sigmod11} propose efficient algorithms for querying probabilistic RDF graphs with node attribute correlations. None of them support 
identity  uncertainty. 

Ioannou et al., \cite{ioannou:pvldb10} propose query evaluation algorithms for uncertain data with identity uncertainty, but their methods are not designed to handle graph data. Furthermore, our semantics are more general, as we allow merge functions to be controlled by the user.
Hua et al., \cite{hua:edbt12} propose a method for evaluating aggregate queries over data with identity uncertainty, but their methods are not designed for graph data either, and their model constrains the acceptable configurations of groups of references representing entities. Our PGM-based representation allows for arbitrary configurations. Dedupalog \cite{arasu:icde09} is a system for declaratively resolving duplicate references using hard and soft constraints. {\sc GrDB} \cite{moustafa:gdm11} is a system for declarative cleaning of noisy graph data, including missing attributes and links, and resolving duplicate references. Neither of these consider the problem of querying uncertain graph data. 

Subgraph pattern matching has received renewed interest in recent years, leading to new 
exact or approximate methods that search for patterns in graph databases consisting either of several relatively small graphs 
or  a single large graph, 
e.g.,~\cite{shasha:pods02,yan:sigmod04,he:icde06,zhao:vldb07,
cheng:sigmod07,
he:sigmod08, zhang:edbt09, zhao:vldb10,fan:vldb10,fan:icde11,zou:vldb09}.
For path indexing, Zhao et al., \cite{zhao:vldb10} use shortest path-based subgraph pattern matching.
As they use certain graphs, issues of combining different types of uncertainty with entity-level semantics do not come up. Further, while we use \emph{context-aware path indexing}, they utilize \emph{shortest paths} calculated at query runtime to prune candidates. Although their use of shortest paths for subgraph pattern matching implies decomposing the query graph into paths as we do, they use different criteria for path decomposition and join order selection better suited for certain graphs. Our approaches utilize probabilistic information for pruning, and implement reduction by join-candidates to further reduce the search space. 
 GraphGrep \cite{shasha:pods02} uses path indexing for querying a database of multiple graphs. It does not handle probabilistic graphs, and it is designed to deal with small graph sizes in the order of tens to hundreds of nodes. For indexing, it indexes paths only without local information. Our approach can be used to query very large probabilistic graphs in the order of millions of nodes and edges.
\newpage
\section{Conclusions and Future Work}
In this paper, we presented a probabilistic approach for modeling uncertain graphs and answering queries over them. Our graph model, probabilistic entity graphs,  captures node attribute uncertainty, edge existence uncertainty, and identity uncertainty. We presented efficient algorithms to solve subgraph pattern matching queries over such uncertain graphs, where queries are expressed and evaluated at  the entity-level. We showed that our approaches outperform an equivalent SQL implementation by multiple orders of magnitude. Future work involves generalizing the graph model to capture other types of entity merging constraints such as transitive closure, and handle other types of uncertainty such as correlations. 


{
\small
\bibliographystyle{abbrv}
\bibliography{references}  
}
\newpage


\end{document}